\documentclass[prd,notitlepage,nofootinbib,superscriptaddress]{revtex4-1}
\usepackage{amsmath}
\usepackage{amsfonts}
\usepackage[utf8]{inputenc}
\usepackage{graphicx}
\usepackage{hyperref}
\usepackage{slashed}
\usepackage{subcaption}
\usepackage{mathrsfs}
\usepackage{array}
\usepackage{amsmath,empheq}
\usepackage{booktabs}
\usepackage{multirow}
\renewcommand{\arraystretch}{1.5}
\newcolumntype{P}[1]{>{\centering\arraybackslash}p{#1}}
\hypersetup{colorlinks, linkcolor = [rgb]{0,0.0,0.75}, citecolor = [rgb]{0,0.0,0.75}, urlcolor = [rgb]{0,0.0,0.75}}
\newcommand{\pvec}[1]{\vec{#1}\mkern2mu\vphantom{#1}}
\DeclareMathOperator{\Tr}{Tr}

\DeclareCaptionJustification{justified}{\leftskip=0pt \rightskip=0pt \parfillskip=0pt plus 1fil}
\makeatletter
\g@addto@macro\bfseries{\boldmath}
\makeatother
\begin{document}

\title{Computing the nucleon charge and axial radii directly at $Q^2=0$ in lattice QCD}

\author{Nesreen~Hasan}
\email{n.hasan@fz-juelich.de}
  \affiliation{Bergische Universität Wuppertal, 42119 Wuppertal, Germany}
  \affiliation{IAS, Jülich Supercomputing Centre, Forschungszentrum Jülich, 52425 Jülich, Germany}

\author{Jeremy~Green}
\email{jeremy.green@desy.de}
  \affiliation{NIC, Deutsches Elektronen-Synchroton, 15738 Zeuthen, Germany}

\author{Stefan~Meinel}
\email{smeinel@email.arizona.edu}
  \affiliation{Department of Physics, University of Arizona, Tucson, AZ 85721, USA}
  \affiliation{RIKEN BNL Research Center, Brookhaven National Laboratory, Upton, NY 11973, USA}

\author{Michael~Engelhardt}
  \affiliation{Department of Physics, New Mexico State University, Las Cruces, NM 88003-8001, USA}

\author{Stefan~Krieg}
  \affiliation{Bergische Universität Wuppertal, 42119 Wuppertal, Germany}
  \affiliation{IAS, Jülich Supercomputing Centre, Forschungszentrum Jülich, 52425 Jülich, Germany}

\author{John~Negele}
  \affiliation{Center for Theoretical Physics, Massachusetts Institute of Technology, Cambridge, MA 02139, USA}

\author{Andrew~Pochinsky}
  \affiliation{Center for Theoretical Physics, Massachusetts Institute of Technology, Cambridge, MA 02139, USA}

\author{Sergey~Syritsyn} 
  \affiliation{RIKEN BNL Research Center, Brookhaven National Laboratory, Upton, NY 11973, USA}
   \affiliation{Department of Physics and Astronomy, Stony Brook University, Stony Brook, NY 11794, USA}

%\author{\dots ?}

\date{\today}
\begin{abstract}
We describe a procedure for extracting momentum derivatives of nucleon matrix elements on the lattice directly at $Q^2=0$. This is based on the Rome method for computing momentum derivatives of quark propagators. We apply this procedure to extract the nucleon isovector magnetic moment and charge radius as well as the isovector induced pseudoscalar form 
factor at $Q^2=0$ and the axial radius. For comparison, we also determine these quantities with the traditional approach of computing the corresponding form factors, i.e. $G^v_E(Q^2)$ and $G_M^v(Q^2)$ for the case of the vector current and $G_P^v(Q^2)$ and $G_A^v(Q^2)$ for the axial current, at multiple $Q^2$ values followed by $z$-expansion fits. We perform our calculations at the physical pion mass using a 2HEX-smeared Wilson-clover action. To control the effects of excited-state contamination, the calculations were done at three source-sink separations and the summation method was used. The derivative method produces results consistent with those from the traditional approach but with larger statistical uncertainties especially for the isovector charge and axial radii.
\end{abstract}

\maketitle

%-------------------
\section{Introduction}
%---------------------
The experimental determinations of the proton (electric) charge radius $r_E^p$ have a discrepancy greater than $5$-sigma between the value determined from spectroscopy of muonic hydrogen~\cite{Pohl2010,Antognini417} and the CODATA average~\cite{Mohr:2015ccw} of experimental results obtained from hydrogen spectroscopy and electron-proton scattering. This presently unresolved ``proton radius puzzle" is the focus of various theoretical and experimental efforts\footnote[1]{Interestingly, a recent result of the proton charge radius obtained by  Beyer et al. using spectroscopic measurements of regular hydrogen has been found to be consistent with the result of the muonic hydrogen experiment~\cite{Beyer79}.}. Last year, the CREMA collaboration reported on their study of muonic
deuterium~\cite{Pohl669}. Their experiment corroborates the muonic hydrogen result for the proton charge radius, while finding a similar $6$-sigma discrepancy for the deuteron charge radius with the
CODATA values, and a $3.5$-sigma discrepancy to electronic deuterium spectroscopy results~\cite{Pohl:2016glp}. 
Thus, having a reliable ab-initio calculation of the proton charge radius is a highly attractive goal for practitioners of lattice QCD.

The conventional approach for determining quantities like the charge radius on the lattice involves the computation of form factors at several different discrete values of the initial and final momenta, $ \vec p$ and $\pvec {p}'$, that are allowed by the periodic boundary conditions, followed by a large extrapolation to zero momentum transfer $Q^2=0$. This introduces a source of systematic uncertainty, analogous to the systematic uncertainty associated with the choices of the fit ansatz and range of $Q^2$ in extracting the proton charge radius from electron-proton scattering data. Systematic errors of this kind have in fact been proposed as a possible explanation of the radius puzzle~\cite{Lorenz:2014vha, Griffioen:2015hta, Higinbotham:2015rja}. Given that the smallest nonzero value of $Q^2$ accessible on the largest available lattices is still an order of magnitude higher than in scattering experiments~\cite{Bernauer:2010wm}, a lattice method for computing $r_E^p$ and similar observables directly at $Q^2=0$ without the need of a shape fit is highly desirable.
%can be used to compute the form factors directly at zero momentum. In essence, this method presents 

The Rome method, presented in Ref.~\cite{deDivitiis:2012vs}, provides a way to calculate the momentum derivatives of quark propagators on the lattice at zero momentum. This enables calculating the momentum derivatives of the correlation functions at zero momentum and obtaining the form factors and their momentum derivatives at vanishing momenta. To this end, one introduces twisted boundary conditions and takes the symbolic derivative(s) with respect to the twist angle (at zero twist angle) before the numerical evaluation of the path integral over the gauge fields.

For the case of a pion, it was shown in Ref.~\cite{Tiburzi:2014yra} that the Rome method for momentum derivatives could be used to calculate the pion charge radius with finite-volume effects that are exponentially suppressed, with asymptotic behavior $\sim \sqrt{m_\pi L} \;e^{-m_\pi L}.$

We employ the Rome method for extracting the proton isovector charge radius $(r_E^2)^v$ and the isovector magnetic moment $\mu^v=G_M^v(0)$, from matrix elements of the vector current. We also extract the proton axial radius $r_A^2$ and the induced pseudoscalar form factor at zero momentum, $G_P(0)$, using nucleon matrix elements of the axial current. We compare the results from the derivative method with those from the traditional approach.
%We compare the derived results on the previously mentioned quantities using the Rome method  to their values using the conventional approach. 

The outline of the paper is as follows. We start by reviewing the electromagnetic and axial form factors in Sec.~\ref{sec1:def-ff}. Section~\ref{sec:2} is devoted  to describing the traditional approach for isolating the nucleon ground state, extracting the nucleon electromagnetic and axial form factors and the fits to the $Q^2$-dependence of the form factors using the $z$ expansion to determine the corresponding radii and form factors at $Q^2=0$. Section~\ref{sec3:DM} explains in detail the derivative method for computing the momentum derivatives of matrix elements at $Q^2=0$ using the Rome method, which we use to determine the charge and axial radii in addition to the magnetic and induced pseudoscalar form factors directly at $Q^2=0$. In Sec.~\ref{LC}, we describe the lattice methodology and the ensemble of configurations that are used. Finally in Sec.~\ref{sec5:results}, we present our numerical results computed directly at $Q^2=0$, and compare them with the traditional approach. We give our conclusions in Sec.~\ref{sec6:conclusion}.
\section{Definitions of the form factors}\label{sec1:def-ff}
The nucleon matrix elements can be parametrized in terms of nucleon form factors as
 \begin{equation}
 \left \langle \pvec{p}',\lambda'|\mathscr{O}_X^{q,\mu} | \vec p,\lambda\right\rangle = \bar u(  \pvec{p}',\lambda')  \mathscr{F}_X^{q,\mu}(\vec p, \pvec{p}') u(\vec p, \lambda),
 \end{equation}
where $\vec{p}$, $\pvec{p}'$ are the initial and final nucleon momenta, $\lambda,\lambda'$ label the different polarization states, and $u$ is the nucleon spinor. We are defining the form factors using a current of flavor $q$ in a proton and $|\vec p,\lambda \rangle$ is a proton state.
$\mathscr{O}_{X}^{q,\mu}$ refers to either the vector ($X=V$) or the axial ($X=A$) current.

For the case of the vector current, $ \mathscr{O}_V^{q,\mu} = \bar q \gamma^\mu q$,  $\mathscr{F}_{V}^{q,\mu}(\vec p, \pvec{p}')$ can be written in terms of the Dirac and Pauli form factors, $F_1^q(Q^2)$ and $F_2^q(Q^2)$, in Minkowski space as:
\begin{equation}
 \mathscr{F}_{V}^{q,\mu}(\vec p, \pvec{p}') = \gamma^\mu F_1^q(Q^2) + \frac{i\sigma^{\mu\nu} (p'-p)_\nu}{2m} F_2^q(Q^2),
\end{equation}
where $m$ is the nucleon mass and $Q^2=-(p'-p)^2\geq 0$ is the momentum transfer. These form factors can also be expressed in terms of the nucleon electric $G_E(Q^2)$ and magnetic  $G_M(Q^2)$ Sachs form factors via:
\begin{align}
G_E(Q^2) &= F_1(Q^2) - \frac{Q^2}{4m^2} F_2(Q^2),\\
G_M(Q^2) &= F_1(Q^2) + F_2(Q^2).
\end{align}
The charge and magnetic radii, $r_{E,M}^2$, and the magnetic moment, $\mu$, are defined from the behavior of $G_{E,M}(Q^2)$ near $Q^2=0$:
\begin{align}
G_E^q (Q^2)& = 1-\frac{1}{6} (r_E^2)^q Q^2 + O(Q^4),\\
G_M^q(Q^2) &= \mu^q \left( 1-\frac{1}{6} (r_M^2)^q Q^2 + O(Q^4)\right).
\end{align}  

For the axial vector current, $\mathscr{O}_A^{q,\mu} = \bar q \gamma^\mu \gamma_5 q$, $\mathscr{F}_{A}^{q,\mu}(\vec p, \pvec{p}')$
can be expressed in terms of the axial and induced pseudoscalar form factors, $G_A^q(Q^2)$ and $G_P^q(Q^2)$, as:
\begin{equation}
\mathscr{F}_{A}^{q,\mu}(\vec p, \pvec{p}') =\gamma^\mu\gamma_5 G_A^q(Q^2) +\gamma_5 \frac{(p'-p)^\mu}{2m} G_P^q(Q^2).
\end{equation}
The axial form factor admits the following expansion for small momentum transfer
\begin{equation}
G_A^q(Q^2) = g_A^q\left(1-\frac{1}{6}(r_A^2)^q Q^2 + O(Q^4) \right),
\end{equation}
where $g_A^q$ is the axial-vector coupling constant and $r_A^q$ is the axial radius. 

In this work, we are considering the isovector electromagnetic Sachs form factors which parametrize the matrix elements of the $u-d$ flavor combination between proton states and, neglecting the isospin breaking effects, are equivalent to the difference between the form factors of the electromagnetic current $V_{em}^\mu=\frac{2}{3} \bar u \gamma^\mu u-\frac{1}{3} \bar d \gamma^\mu d $ in a proton and in a neutron, $G_{E,M}^{p,n}(Q^2)$,
%the proton and neutron form factors and defined as:
\begin{equation}
G_{E,M}^v (Q^2) = G_{E,M}^p(Q^2) - G_{E,M}^n (Q^2)=G_{E,M}^u(Q^2) - G_{E,M}^d (Q^2) \equiv G_{E,M}^{u-d}(Q^2).
\end{equation}
The isovector axial form factors $G_{A,P}^v(Q^2)$ are defined in a similar way.
 %------------------------
 \section{Computation of matrix elements using the traditional method}
  \label{sec:2}
 %----------------
 For determining the nucleon matrix elements in lattice QCD, we compute the nucleon two-point and three-point functions,
 \begin{equation}\label{defc2}
 C_{2}(\vec p, t) = \sum_{\vec x} e^{-i\vec p \vec x} \sum_{\alpha\beta} \left[(\Gamma_{\text{pol}})_{\alpha\beta} \left\langle \chi_\beta(\vec x,t) \bar{\chi}_\alpha(0)\right\rangle\right],
 \end{equation}
 \begin{equation} \label{defc3}
 C_{3}^{ \mathscr{O}_X^{q,\mu}}(\vec p,\pvec{p}', \tau, T) = \sum_{\vec x, \vec y} e^{-i \pvec{p}' \vec x} e^{i( \pvec{p}' - \vec p)\vec y} \sum_{\alpha\beta}\left[(\Gamma_ {\text{pol}})_{\alpha\beta} \left\langle \chi_{\beta}(\vec x, T) \mathscr{O}_X^{q,\mu}(\vec y, \tau) \bar \chi_{\alpha}(0)\right\rangle\right].
 \end{equation}
In this section, we use Minkowski-space gamma matrices. Above, $\chi=\epsilon^{abc} (\tilde{u}_a^T C \gamma_5 \frac{1+\gamma_0}{2} \tilde{d}_b) \tilde{u}_c$ is a proton interpolating operator constructed using smeared quark fields $\tilde q$ and $\Gamma_{\text{pol}} = \frac{1}{2}(1+\gamma_0)(1+\gamma_3\gamma_5)$ is a spin and parity projection matrix.
%We compute $C_{3}$ with both $ \pvec{p}'= \vec 0$ and $ \pvec{p}' = \frac{2\pi}{L} (-1,0,0)$, and for the quark flavors $q\in \{u,d\}$.
 The three-point correlators have contributions from both connected and disconnected quark contractions, but we compute only the connected part since  for the isovector flavor combination the disconnected contributions cancel out.
 
We will be tracing the correlators with  $\Gamma_{\text{pol}}$ which contains the projector $(1+\gamma_0)/2$ so that we can effectively write the overlap of the interpolating operator with the ground-state proton as $\langle \Omega | \chi_\alpha(0) | \vec p,\lambda\rangle = {Z(\vec p)} u(\vec p, \lambda)_\alpha$~\cite{Bowler:1997ej,Capitani:2015sba}. At large time separations we obtain
 \begin{align}
 C_{2}(\vec p, t) &= \frac{Z(\vec p)^2 e^{-E(\vec p)t}}{2E(\vec p)} \Tr[\Gamma_{\text{pol}}(m + \slashed p)]\left( 1 + O(e^{-\Delta E_{10}(\vec p)t})\right),\label{eq:gt-c2}\\
C_{3}^{\mathscr{O}_X^{q,\mu}} (\vec p, \pvec {p}',\tau,T) &= \frac{{Z(\vec p)Z(\pvec {p}')} e^{-E(\vec p) \tau - E(\pvec {p}') (T-\tau)} }{4E(\pvec {p}')E(\vec p)} \sum_{\lambda,\lambda'} \bar u(\vec p, \lambda) \Gamma_{\text{pol}} u(\pvec{p}', \lambda') \langle p',\lambda' | \mathscr{O}_X^{q,\mu} | p,\lambda \rangle \label{eq:gt-c3}\\
&\qquad \qquad \qquad \qquad \times \left( 1+ O(e^{-\Delta E_{10}(\vec p) \tau}) + O(e^{-\Delta E_{10}(\pvec{p}')(T-\tau)})\right)\nonumber,
 \end{align} 
 where $\Delta E_{10}(\vec p )$ is the energy gap between the ground and the lowest excited state with momentum $\vec p$. By taking $\tau$ and $T-\tau$ to be large,  unwanted contributions from excited states can be eliminated.
 In order to compute $C_ {3}$, we use sequential propagators through the sink~\cite{Martinelli:1988rr}. This has the advantage of allowing for any operator to be inserted at any time using a fixed set of quark propagators, but new backward  propagators must be computed for each source-sink separation $T$. Increasing $T$ suppresses excited-state contamination, but it also increases the noise; the signal-to-noise ratio is expected to decay asymptotically as $e^{-(E -\frac{3}{2} m_\pi) T}$~\cite{Lepage:1989hd}.

In order to cancel the overlap factors and the dependence on Euclidean time, we define the \emph{normalization 
ratio}, $R_N^X$, and the \emph{asymmetry ratio}, $R_S$ as
 \begin{align}
 R_N^X &=\frac{C_{3}^{\mathscr{O}_X^{q,\mu}} (\vec p, \pvec {p}', \tau,T)} {\sqrt{C_{2} (\vec p, T) C_{2}(\pvec{p}', T) }}, \label{eq:R_N}\\
 R_{S} &=  \sqrt{\frac{C_{2}(\vec p, T-\tau) C_{2}(\pvec {p}',\tau)}{C_{2}(\pvec {p}', T-\tau) C_{2}(\vec p,\tau)}},\label{eq:R_S}
 \end{align}
and compute their product
 \begin{align}
 R_X^{q,\mu}(\vec p, \pvec{p}', \tau,T) =R_N^X R_S =  M_X^{q,\mu}(\vec p, \pvec{p}') + O(e^{-\Delta E_{10}(\vec p) \tau}) + O(e^{-\Delta E_{10}(\pvec{p}')(T-\tau)}) + O(e^{-\Delta E_{min}T}), \label{rnra}
 \end{align}% 
 as a function of $\tau \in [0,T]$ with fixed $T$. Above,
\begin{equation}
M_X^{q,\mu}(\vec p, \pvec{p}') = \frac{ \sum_{\lambda,\lambda'} \bar u(\vec p, \lambda) \Gamma_{\text{pol}} u(\pvec{p}', \lambda') \langle p',\lambda' | \mathscr{O}_X^{q,\mu} | p,\lambda \rangle}{4 \sqrt{E(\vec p) E(\pvec{p}') (E(\vec p) +m)(E(\pvec{p}')+m) }}.
\end{equation} 
and $\Delta E_{\text{min}} = \min\{ \Delta E_{10}(\vec p) , \Delta E_{10}(\pvec {p}')\}$.

%The \emph{normalization ratio}, $R_N^X$, and the $asymmetry\;ratio, R_S$, are defined as:

 The ratio in Eq.~\eqref{rnra} gives an estimate of the nucleon matrix element $\langle p',\lambda' | \mathscr{O}_X^{q,\mu} | p,\lambda \rangle$ and produces at large $T$ a plateau with ``tails'' at both ends caused by excited states. In practice, for each fixed $T$, we average over the central two or three points near $\tau=T/2$, which allows for matrix elements to be computed with errors that decay asymptotically as $e^{-\Delta E_{\text{min}}T/2}$. 
 
 Improved asymptotic behavior of excited-state contributions can be achieved by using the \emph{summation method}~\cite{Capitani:2010sg,Bulava:2010ej} which requires performing the calculations with multiple source-sink separations. Taking the sums of ratios for each $T$ yields
 \begin{equation}
 S_X^{q,\mu}(\vec p, \pvec{p}',T) \equiv \sum_{\tau=\tau_0}^{T-\tau_0} R_X^{q,\mu}(\vec p, \pvec{p}',\tau,T) = c+T M_X^{q,\mu}(\vec p, \pvec{p}') + O(Te^{-\Delta E_{\text{min}} T}),
 \end{equation}
 where we choose $\tau_0=1$ and $c$ is an unknown constant. The matrix element can then be extracted from the slope of a linear fit to  $S_X^{q,\mu}(\vec p, \pvec{p}', T)$ at several values of $T$. The leading excited-state contaminations decay now as $Te^{-\Delta E_{min} T}$.

For calculating the form factors --- $G_E(Q^2), G_M(Q^2)$ for the case of the vector current and  $G_A(Q^2), G_P(Q^2)$ for the case of the axial current --- we construct a system of equations parameterizing the corresponding set of matrix elements at each fixed value of $Q^2$~\cite{Hagler:2003jd}. We combine equivalent matrix elements to improve the condition  number~\cite{Syritsyn:2009mx}.  We find the solution of the resulting overdetermined system of equations by performing a linear fit. This approach makes use of all available matrix elements in order to minimize the statistical uncertainty in the resulting form factors.
%--------------
%\subsection{Form factor fits using the z expansion}

The charge and axial radii can be extracted from the slopes of the electric and axial form factors at $Q^2=0$, respectively.
For that we need to fit the $Q^2$-dependence of each form factor.
 In order to avoid the model dependence included in the commonly used fit ansatzes,
 such as a dipole, we use the model-independent $z$ expansion~\cite{Bhattacharya:2011ah,Bhattacharya:2015mpa,Hill:2010yb,Epstein:2014zua},
 where each form factor can be described by a convergent Taylor series in $z$
\begin{equation}
G(Q^2) = \sum_k^{k_{\text{max}}} a_k z^k,\qquad z=\frac{\sqrt{t_{\text{cut}} + Q^2} - \sqrt{t_{\text{cut}}}}{\sqrt{t_{\text{cut}} + Q^2} + \sqrt{t_{\text{cut}}}},
\end{equation}
which conformally maps the complex domain of analyticity in $Q^2$ to $|z|<1$. We fix $a_0=1$ for fitting $G_E(Q^2)$ since $G_E(0)=1$. We use the particle production threshold $t_{\text{cut}} = (2m_\pi)^2$ for the vector case and $t_{\text{cut}} = (3m_\pi)^2$ for the axial case. We apply $z$-expansion fits 
following the approach of Ref.~\cite{Green:2017keo}.
The intercept and slope of the form factor at $Q^2=0$ can be obtained from the first two coefficients, $a_0$ and $a_1$. We impose Gaussian priors on the remaining coefficients centered at zero with width equal to $5 \text{max}\{|a_0|,|a_1|\}$.
We truncate the series with $k_{\text{max}} = 5$ after verifying that using a larger $k_{\text{max}}$ produces identical fit results in our probed range of $Q^2$.

Furthermore, the isovector $G_P$ form factor has an isolated pole  at the pion mass below the particle production threshold. We thus remove this pole before fitting and perform the $z$-expansion fit to $(Q^2 + m_\pi^2) G_P(Q^2)$.

We perform correlated fits by minimizing
\begin{equation}
\chi^2_{\text{aug}} = \sum_{i,j} \left( G(Q_i^2) - \sum_k a_k z(Q_i^2)^k\right) S_{ij}^{-1} \left( G(Q_j^2) - \sum_{k'} a_{k'} z(Q_j^2)^{k'}\right) + \sum_{k>1} \frac{a_k^2}{w^2},
\end{equation}
with respect to $\{a_k\}$, where $S$ is an estimator of the covariance  matrix and the last term augments the chi-squared with the Gaussian priors. For choosing the estimator of the covariance matrix, we  use $S=(1-\lambda)C + \lambda C_{\text{diag}}$, where $\lambda=0.1$, $C$ is the bootstrap estimate of the covariance matrix and $C_{\text{diag}}$ is the diagonal part of $C$.
 %-----------
\section{Derivative method}\label{sec3:DM}
%----------
In this section, we explain the details of our approach for extracting the nucleon charge radius directly at $Q^2=0$. We begin with reviewing the Rome method for computing the momentum derivatives of quark propagators in Subsection~\ref{Romemethod}. The flavor structure of the correlators constructed from the momentum derivatives of the quark propagators is investigated in~\ref{flavourstructure}.
In Subsection~\ref{sec:derivative_2pt_3pt}, we show how to use the momentum derivatives of the quark propagator in order to obtain the first- and second-order derivatives of the nucleon two- and three-point functions with respect to the initial-state momentum $\vec p$, and then obtain momentum derivatives of matrix elements in Subsection~\ref{derivative_of_ratio}. From the latter one can then extract the charge radius $r_E^2$, the magnetic moment $\mu = G_M(0)$, for the case of the electromagnetic vector current, and the axial radius, $r_A^2$, and the induced pseudoscalar form factor at zero momentum, $G_P(0)$, for the case of the axial current.
%----
\subsection{Momentum derivatives of quark propagator} \label{Romemethod}

On a lattice with finite size and quark fields satisfying periodic boundary conditions, consider a generic correlation function $C(\vec p,t)$ depending on the three-momentum $\vec p$ and Euclidean time $t$, which after fermionic integration and Wick contractions can be written in terms of quark propagators and operator insertions as, 
\begin{equation}
C(\vec p,t) = \int dU P[U] \sum_{\vec x,\dots} e^{-i\vec p(\vec x-\vec y)} \Tr\{ G[x,y;U] \Gamma \dots \},   
 \label{eq:G}
\end{equation}
where $U$ are gauge links and $P[U]$ is the corresponding probabilistic weight in the functional integral. The plane-wave phase factor $e^{-i\vec p(\vec x-\vec y)}$ can then be absorbed into one of the quark propagators, which results in a momentum dependent quark propagator $G[x,y;U, \vec p] = e^{-i\vec p(\vec x-\vec y)}G[x,y;U]$.
$G[x,y;U, \vec p]$ can be obtained by solving the lattice Dirac equation with link variables rescaled by a phase factor:
\begin{align}
 &U_k(x) \to e^{ip_k} U_k(x),\\
 &\sum_y D[x,y;U, \vec p]G[y,z;U, \vec p] =  \delta_{x,z}.
\end{align}
Carrying momentum in a propagator with a uniform $U(1)$ background 
field is the same approach as used in a standard transformation of twisted 
boundary conditions~\cite{Bedaque:2004kc,deDivitiis:2004kq}. With $\vec p$ restricted to be a Fourier momentum 
in the finite volume, the above redefinition is exact. However, to 
obtain a momentum derivative, we must implicitly make use of twisted 
boundary conditions and allow $\vec p$ to be continuous. We use the expansion 
of the lattice Dirac operator
\begin{equation}
D[U, \vec p] = D[U] + p_k \frac{\partial  D}{\partial p_k} \Big |_{\vec p=\vec 0} + \frac{p_k^2}{2} \frac{\partial^2 D}{\partial p_k^2} \Big |_{\vec p = \vec 0} + \dots ,
\end{equation}
and $D[U,\vec p] G[U,\vec p] = 1$ to compute the first-order momentum derivative of the propagator as:
\begin{equation}\label{eq:tmp}
\frac{\partial D}{\partial p_k} G + D \frac{\partial G}{\partial p_k} = 0,
\end{equation}
where we use the compact notation
\begin{equation}
\frac{\partial D}{\partial p_k} \equiv \frac{\partial D[\dots; U,\vec p]}{ \partial p_k} \Big |_{\vec p=0},
\end{equation}
and similar notation for $G(\dots;U,\vec p)$. Multiplying Eq.~\eqref{eq:tmp} from the left by $G \equiv D^{-1}$ leads to:
\begin{equation}
\frac{\partial G}{\partial p_k} = -G \frac{\partial D}{\partial p_k} G.
\end{equation}
Similarly, we can derive the second-order momentum derivative of the propagator:
\begin{equation}
\frac{1}{2} \frac{\partial^2 G}{\partial p_k^2} = +G \frac{\partial D}{\partial p_k} G \frac{\partial D}{\partial p_k} G - G\frac{1}{2} \frac{\partial^2 D}{\partial p_k^2} G.
\end{equation}
Using the lattice Dirac operator for the clover-improved Wilson action, the momentum derivatives of the propagators at a fixed gauge background become~\cite{deDivitiis:2012vs}:
\begin{align}
\frac{\partial}{\partial p_k} G(x,y; \vec p) \Big |_{\vec p= \vec 0} &= -i\sum_z G(x,y)  \Gamma_V^k G(x,y),\label{prop_first_der} \\
\frac{\partial^2}{\partial p_k^2} G(x,y; \vec p) \Big |_{\vec p= \vec 0} &= -2\sum_{z,z'} G(x,z) \Gamma_V^k G(z,z') \Gamma_V^k G(z',y)   -\sum_z G(x,y) \Gamma_T^k G(x,y). \label{prop_second_der}
\end{align}
We drop $U$ from the propagators for brevity.
 $\Gamma_V^k$ and $\Gamma_T^k$ are the point split vector and tadpole currents, respectively. Those are defined using Euclidean gamma matrices, $\gamma^k_E$, as:
\begin{align}
\Gamma_V^k G(z,y;U) &\equiv U_j^\dagger(z-\hat k) \frac{1+\gamma_E^k}{2} G(z-\hat k,y) - U_k(z) \frac{1-\gamma_E^k}{2} G(z+\hat k,y), \\
\Gamma_T^k G(z,y;U) &\equiv U_j^\dagger(z-\hat k) \frac{1+\gamma_E^k}{2} G(z-\hat k,y) + U_k(z) \frac{1-\gamma_E^k}{2} G(z+\hat k,y).
\end{align}

In the case of a smeared-source smeared-sink propagator (needed in the two-point function), the phase factor can be absorbed into the propagator in the following way:
\begin{align}
\tilde{\tilde{G}}(x,y;\vec p) &= e^{-i{\vec p(\vec x-\vec y)}} \sum_{x',y'} K(x,x') G(x',y') K(y',y) \nonumber \\
&=\sum_{x',y'} \underbrace{e^{-i{\vec p(\vec x-\pvec{x}')}} K(x,x')}_{ K(x,x';\vec p)}  \underbrace{e^{-i{\vec p(\pvec{x}'-\pvec{y}')}} G(x',y')}_{G(x',y';\vec p)}  \underbrace{e^{-i{\vec p(\pvec{y}'-\vec y)}} K(y',y)}_{ K(y',y;\vec p)} ,
%&= \sum_{x',y'} K(x,x';\mathbf p) G(x',y';\mathbf p) K(y',y;\mathbf p),
\end{align}
\normalsize
where $K$ is the smearing kernel. The momentum derivatives can then be calculated using the product rule along with Eq.~\eqref{prop_first_der} and Eq.~\eqref{prop_second_der}. Denoting the momentum derivative with $'$ for shorter notation, we obtain
\begin{align}
(KGK)' &= K'GK + K(GK)' ,\label{eq:1KGK} \\
(KGK)'' &= K'' GK +2K'(GK)'+K(GK)''.\label{eq:2KGK}
\end{align}
For the smeared-source point-sink propagator, which is needed for the three-point function and for evaluating Eq.~\eqref{eq:1KGK} and Eq.~\eqref{eq:2KGK}, we obtain
\begin{align}
 (GK)' &= G[-i\Gamma_V G K + K'], \\
 (GK)''&= G[-2i\Gamma_V(GK)' - \Gamma_T GK + K''].
\end{align}
Organized in this way, we have one additional propagator right-hand-side per derivative.
Gaussian Wuppertal smearing~\cite{Gusken:1989qx} is given by \small$K(x,y;\vec p) = \sum_{x',x'',...} \underbrace{K_0(x,x';\vec p) K_0(x',x'';\vec p) ... K_0(x^{'...'},y;\vec p)}_{N_W}$, \normalsize with
%\small
\begin{align}
K_0(x,y;\vec p) &= e^{-i{\vec p(\vec x-\vec y)}} \frac{1}{1+6\alpha} \Bigg(\delta_{x,y} + \alpha\sum_{j=1}^3 \Big[\ \tilde{U}_j(x) \delta_{x+\hat{\jmath},y}  + \tilde{U}_j^\dagger (x-\hat{\jmath}) \delta_{x-\hat{\jmath},y}\Big]\Bigg)\nonumber \\
&= \frac{1}{1+6\alpha} \Bigg( \delta_{x,y} + \alpha \sum_{j=1}^3  \Big[ e^{ip^j} \tilde{U}_j(x) \delta_{x+\hat{\jmath},y} + e^{-ip^j} \tilde{U}_j^\dagger(x-\hat{\jmath}) \delta_{x-\hat{\jmath},y}\Big]\Bigg).
\end{align}
\normalsize
We use APE-smeared gauge links $\tilde{U}$~\cite{Albanese:1987ds}. The $m$th derivative of $K_0$ at zero momentum is equal to
\begin{equation}
K_0^{(m)}(x,y) \equiv \Big( \frac{\partial}{\partial p^j}\Big)^m K_0(x,y;\vec p) \Bigg |_{\vec p=0}= \frac{\alpha}{1+6\alpha} \Bigg[ i^m \tilde{U}_j(x) \delta_{x+\hat{\jmath},y} + (-i)^m \tilde{U}_j^\dagger (x-\hat{\jmath}) \delta_{x-\hat{\jmath},y}\Bigg].
\end{equation}
Thus, the first- and second-order momentum derivatives of smearing with $N_W$ iterations, $K = K_0^{N_W}$, can be computed iteratively using $(K_0^N)' = K_0'K_0^{N-1} + K_0(K_0^{N-1})'$ and $(K_0^N)'' = K_0'' K_0^{N-1} + 2 K_0'(K_0^{N-1})' + K_0(K_0^{N-1})''.$

\subsection{Flavor structure of correlators constructed from propagator derivatives}\label{flavourstructure}
In cases where derivatives of nucleon
two-point functions need to be evaluated, there is an ambiguity in applying
the above procedure: there are three quark propagators, and the momentum could
be absorbed into any of them. To resolve this issue, we make explicit
use of twisted boundary conditions, with the understanding that before
computing any correlation functions we will take the derivative with
respect to the twist angle, at vanishing twist angle.

We introduce a third light quark, $r$, with the same mass as $u$ and
$d$ but with twisted boundary conditions, and a corresponding ghost
quark that cancels its fermion determinant. The three light quarks
$\{u,d,r\}$ contain an approximate SU(3) flavor symmetry that becomes
exact when the twist angle is zero, or in the infinite-volume
limit. Under this symmetry group there is a baryon octet that contains
the ordinary (untwisted) nucleons, as well as states with one or two
$r$ quarks. We are interested in the states with one $r$ quark, and we
find that there are two kinds: an isospin singlet and a triplet, the
$\Lambda_r$ and $\Sigma_r$, respectively. This was previously
discussed in Ref.~\cite{Jiang:2008ja}.

For the states with quark content $udr$ we use interpolating operators
\begin{equation}
\begin{aligned}
  \chi_{\Sigma_r} &= \frac{1}{\sqrt{2}}\left( [rud] + [rdu] \right),\\
  \chi_{\Lambda_r} &= \frac{1}{\sqrt{6}}\left( 2[udr] - [rud] - [dru] \right),
\end{aligned}
\end{equation}
where $[pqr]\equiv\epsilon^{abc}(\tilde p^T_a
C\gamma_5\frac{1+\gamma_0}{2}\tilde q_b)\tilde r_c$.  When contracted
with the projector $\frac{1+\gamma_0}{2}$, the flavor-singlet operator, $1/\sqrt{3}([udr] + [rud] + [dru])$, 
vanishes and the $\Lambda_r$ operator can be simplified to
$\chi_{\Lambda_r} = \sqrt{\frac{3}{2}}[udr]$. We consider three-point
functions for the transition from a state with one $r$ quark to an
ordinary nucleon:
\begin{equation}
  C_3^{X\to N}(\vec p,\pvec{p}',\tau,T) = \sum_{\vec x,\vec y}e^{-i\pvec{p}'(\vec x-\vec y)}\Tr\left[\Gamma_\text{pol}
 \langle \chi(\vec x,T) \mathscr{O}(\vec y,\tau) \bar\chi_X(0)\rangle\right],
\end{equation}
where $\mathscr{O}=\bar u\Gamma r$ is a quark bilinear and $X$ is $\Sigma_r$ or $\Lambda_r$. The initial momentum $\vec p$ is implied in the initial state due to the twisted boundary conditions for the $r$ quark. The ground-state contribution is proportional to the matrix element $\langle N(\pvec{p}')|\mathscr{O}|
X(\vec p)\rangle$ for which we will evaluate $\frac{\partial}{\partial
  \vec p}$ at $\pvec{p}'=\vec p=0$. In practice, we simply use our
already coded expressions for the connected diagrams in the nucleon
three-point functions $C_3^q$ with $\mathscr{O}_q=\bar q\Gamma q$,
$q\in\{u,d\}$, and replace the propagator connecting the nucleon
source and $\mathscr{O}_q$ with a first- or second-derivative
propagator. By comparing the contractions, we find the relations
\begin{equation}\label{eq:ud_LS}
\begin{aligned}
  C_3^{\Sigma_r\to N} &= \frac{1}{\sqrt{2}}C_3^d, \\
  C_3^{\Lambda_r\to N} &= \frac{1}{\sqrt{6}}\left( 2 C_3^u - C_3^d \right),
\end{aligned}
\end{equation}
where the $r$ propagator is substituted into the evaluation of the
right-hand-side expressions as described above. A similar
consideration was made in Ref.~\cite{Jiang:2008ja}; these relations
could also be derived from SU(3) symmetry.

When forming ratios, we must use the appropriate two-point functions:
taking Eq.~\eqref{rnra} with the three-point function $C_3^{X\to N}$,
all nucleon two-point functions that take the initial-state momentum
$\vec p$ must be replaced by the two-point function for state $X$.
Once we have formed the ratios for the $X\to N$ matrix elements, we
can invert the relations in Eq.~\eqref{eq:ud_LS} to obtain the nucleon
matrix elements of $\mathscr{O}_{u}$ and $\mathscr{O}_d$.
%%%%%%%%
\subsection{Momentum derivatives of the two-point and three-point functions}
\label{sec:derivative_2pt_3pt}
\begin{figure}%[H]
\centering
\includegraphics[width=0.3\textwidth]{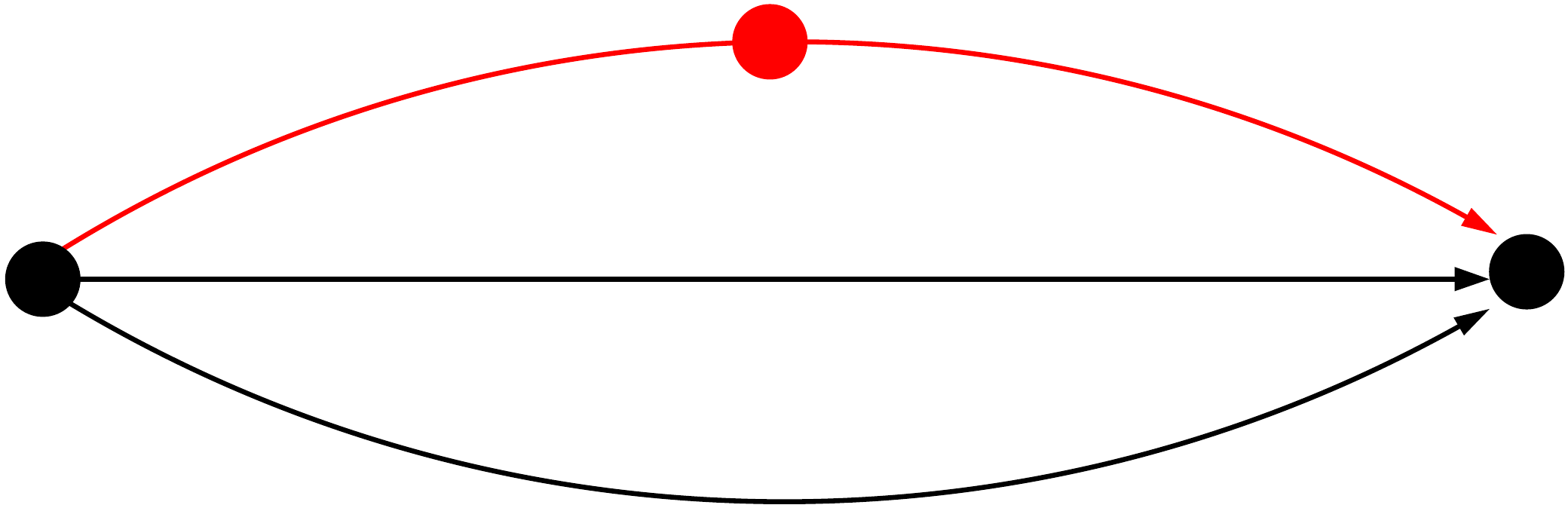}
\includegraphics[width=0.6\textwidth, trim = -1cm 2cm 0cm 0cm]{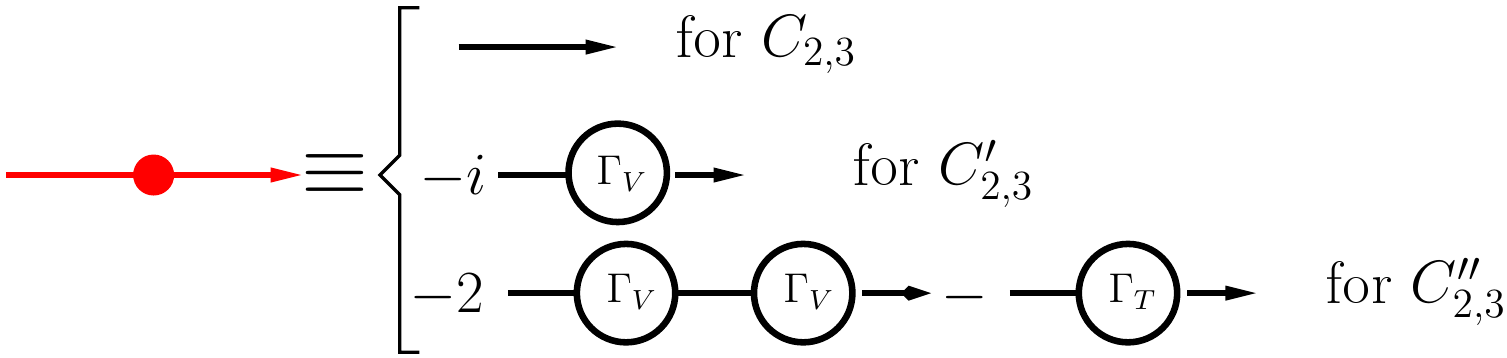}\\
\vspace{0.7cm}
\includegraphics[width=0.3\textwidth]{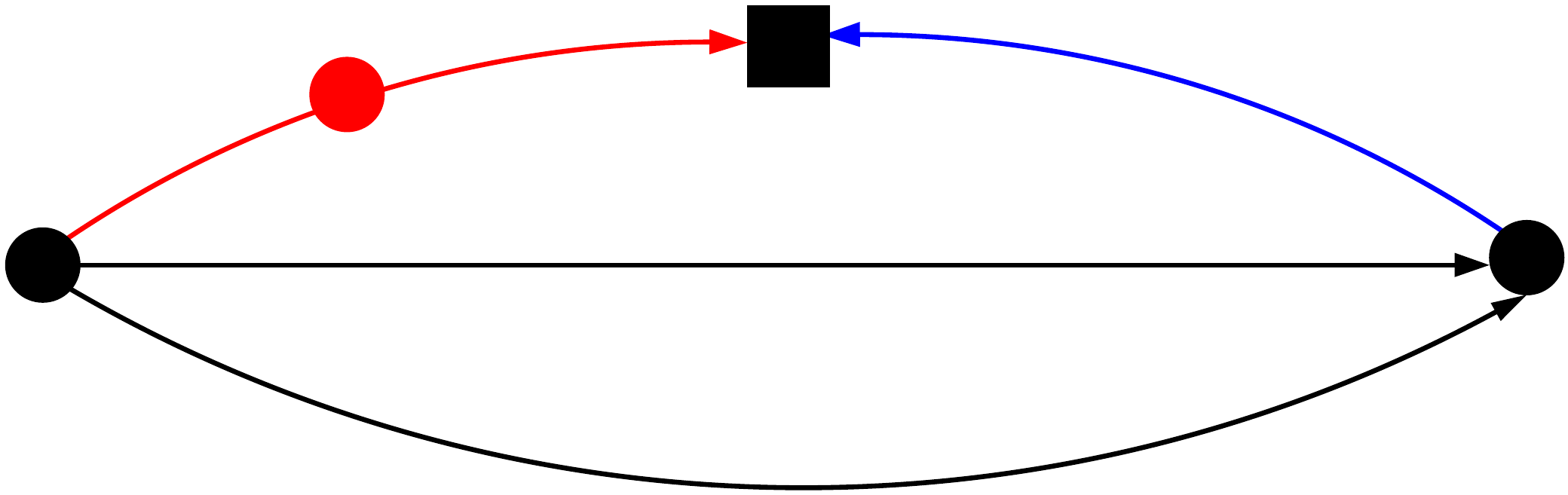}
\hspace{10.5cm}
  \caption{Left: Nucleon two-point (top) and three-point (bottom) functions . The solid black circles represent the nucleon source and sink, the black square in the three-point function represents the current insertion. The red line refers to the propagator which we use for computing the momentum derivatives of the correlators which carry therefore  the derivative vertex (solid red circle). The right panel shows the representation of the derivative vertex for the simplified case of unsmeared propagators.}
  \label{fig:diagrams}
\end{figure}

Let us consider the two-point function of the isospin singlet operator, ${\chi}_{\Lambda_r} = \sqrt{\frac{3}{2}} [udr]$. This can be written in terms of smeared-source smeared-sink quark propagators, $\tilde{\tilde{G}}$, as
\begin{align}
C_{2}^{\Lambda_r}(\vec p, t)&= \frac{3}{2} \sum_{\vec x} e^{-i{\vec p \vec x}} \epsilon^{abc}\epsilon^{def} \sum_{\alpha\beta} (\Gamma_\text{pol})_{\alpha\beta} f_{\beta\gamma\delta\epsilon} \bar f_{\alpha\zeta\eta\theta} \big\langle \tilde{\tilde{G}}_{\gamma\theta}^{af}(x,0) 
\tilde{\tilde{G}}_{\delta\eta}^{be}(x,0)
\tilde{\tilde{G}}_{\epsilon\zeta}^{cd}(x,0)
  \big \rangle\nonumber\\
&= \frac{3}{2}\sum_{\vec x} \epsilon^{abc}\epsilon^{def} \sum_{\alpha\beta} (\Gamma_\text{pol})_{\alpha\beta} f_{\beta\gamma\delta\epsilon} \bar f_{\alpha\zeta\eta\theta} \big \langle \tilde{\tilde{G}}_{\gamma\theta}^{af}(x,0)
 \tilde{\tilde{G}}_{\delta\eta}^{be}(x,0)
 \tilde{\tilde{G}}_{\epsilon\zeta}^{cd}(x,0;\vec {p})
   \big \rangle,
\end{align}
where $f_{\beta\beta\gamma\delta}$ is the spin tensor determining the quantum numbers of the $\Lambda_r$ and $\tilde{\tilde{G}}(x,0;\vec p) = e^{-i{\vec p \vec x}} \tilde{\tilde{G}}(x,0)$. By using the first- and second-order momentum derivatives of a quark propagator at zero momentum given in Eq.~\eqref{prop_first_der} and Eq.~\eqref{prop_second_der}, one can straightforwardly calculate the momentum derivatives of the two-point correlators.

For connected diagrams, the three-point function with current $O_\Gamma=\bar q \Gamma q$ and zero sink momentum $\pvec{p}' =0$ can be written as
\begin{equation}
C_3(\vec p,\tau,T) = \sum_{{\vec x,\vec y}} e^{-i{\vec p \vec y}} \sum_{\alpha\beta}(\Gamma_\text{pol})_{\alpha\beta} \left\langle \chi_\beta (\vec x,T) O_\Gamma (\vec y,\tau) \bar \chi_\alpha(0)\right\rangle  \sim \sum_{\vec y}\left \langle G_S(y) \Gamma \tilde{G}(y,0;\vec p)\right\rangle, 
\end{equation}
where $\tilde{G}$ refers to a propagator with smeared source and point sink and $G_S(y)$ is the sequential backward propagator, which is independent of $\vec p$. 
Only the forward propagator $\tilde{G}(y,0;\vec p)$ needs to be expanded using Eq.~\eqref{prop_first_der} and Eq.~\eqref{prop_second_der}. Hence, no additional backward propagators are needed. Figure~\ref{fig:diagrams} shows graphically the way we compute the momentum derivatives of the  correlation functions on the quark level.
The derivative method cannot be
applied to disconnected diagrams because those involve a quark
propagating from a point to the same point and therefore the momentum transfer can not be absorbed into the propagator.
%---------------------
\subsection{Momentum derivatives of the ratio}
\label{derivative_of_ratio}
%------------------------
Because we do not know how $Z(\vec p)$ depends on the momentum, we need to compute the momentum derivatives of the ratio of three-point and two-point functions given in Eq.~\eqref{rnra}. 
Here and in the following, we use Minkowski-space gamma matrices. We set $\pvec{p}'=0$ and $\vec p = k \vec e_j$, where $\vec e_j$ is the unit vector in $j$-direction. For computing the first- and second-order momentum derivatives of the ratio in Eq.~\eqref{rnra}, we start by computing the momentum derivatives of the normalization ratio part, $R_N^X$, defined in Eq.~\eqref{eq:R_N}:
\begin{align}
\left(R_N^X(k)\right)' &= \frac{-C'_2(k)C_3(k) + 2 C_2(k)C_3'(k)}{2\sqrt{C_2(0) C_2(k)^3}}, \label{fd}\\
\left(R_N^X(k)\right)'' &= \frac { (3[C'_2(k)]^2 - 2C_2(k) C''_2(k))C_3(k) + 4C_2(k) (-C'_2(k)C_3'(k) + C_2(k) C_3''(k) )}{4\sqrt{C_2(0)C_2(k)^5}},\label{sd}
\end{align}
where, for more readability we suppress the $\tau, T$ parameters as well as $\mathcal O_X^\mu$ from the correlation functions and the ratio. We denote the derivatives with a prime, e.g. $C'_2(k) \equiv \frac{dC_2(k)}{dk} $. 
We know that $C'_2(0) = 0$ in the infinite-statistics limit because of parity symmetry. Hence, we can eliminate this from the ratios.
Similarly, we can calculate $R'_S(k)$ and $R''_S(k)$ which can be used together with Eq.~\eqref{fd} and Eq.~\eqref{sd} to calculate the first- and second-order derivatives of the ratio $R_X$. These derivatives are computed on the lattice directly at $k=0$ as discussed earlier in the previous section.

From the ground-state contributions to the correlation functions given in Eq.~\eqref{eq:gt-c2} and Eq.~\eqref{eq:gt-c3}, we find the following ground-state contribution to their ratio 
\begin{equation}
R_X(k) = \frac{\Tr\left[\Gamma_\text{pol}\mathscr F_X(k) (m+E\gamma^0-k\gamma^j)\right]}{2\sqrt{2E(E+m)}}.
\end{equation}
We take the derivative with respect to $k$ and obtain: 
\begin{align}
(R_X)'(k)  &= \frac{\Tr\left[ \Gamma_\text{pol}\left(\mathscr F_X'(k)(m+E\gamma^0-k\gamma^j) + \mathscr F_X(k) (E'\gamma^0-\gamma^j)\right)\right]}{2\sqrt{2E(E+m)}} \\ & \qquad- \frac{\Tr\left[\Gamma_\text{pol} \mathscr F_X(k) (m+E\gamma^0 -k\gamma^j)\right](2E+m)E' }{4\sqrt 2 [E(E+m)]^{3/2}} \nonumber.
\end{align}
$(R_X)''(k)$ can be calculated in a similar way.
We use the continuum dispersion relation $E(k) = \sqrt{m^2+k^2}$, which implies $Q^2 = 2m\sqrt{m^2+k^2} -2m^2$, and find that at $k=0$, the second derivative is needed to obtain the slope of $F_1$:
\begin{equation}
\frac{dF_1}{dk}\Big|_{k=0} = \frac{dQ^2}{dk}\Big|_{k=0}\frac{dF_1}{dQ^2}\Big|_{Q^2=0} = 0, \qquad
 \frac{d^2F_1}{dk^2}\Big|_{k=0} = 2 \frac{dF_1}{dQ^2}\Big|_{Q^2=0}.
\end{equation}
The same applies  for $F_2, G_A,$ and $G_P$.
Furthermore, we have at $k=0$ :%$E(0) =m,\; E'(0) = 0, \; E''(0) = 1/m$ and 
\begin{equation}
E(0) =m,\qquad E'(0) = 0, \qquad E''(0) = 1/m,
\end{equation}
\begin{equation}
\mathscr F_V(0) = F_1(0)\gamma^\mu, \quad \mathscr F_V'(0) = F_2(0)\frac{i\sigma^{\mu j}}{2m},  \quad   \mathscr F_V''(0) = 2\frac{d F_1}{dQ^2}\Big|_{Q^2=0} \gamma^\mu - F_2(0) \frac{i\sigma^{\mu0}}{2m^2},
\end{equation}
\begin{align}
&\mathscr F_A(0) = G_A(0)\gamma^\mu \gamma_5,\quad  \mathscr F_A'(0)=\begin{cases}
    -\frac{1}{2m} G_P(0)\gamma_5, & \mu=j\\
    0, & \mu \neq j
  \end{cases},\\
  & \mathscr F_A''(0)= 2\frac{d}{dQ^2}G_A(0)\gamma^\mu\gamma_5 + \begin{cases}
  -\frac{1}{2m^2} G_P(0)\gamma_5, &\mu=0\\
  0, &\mu \neq 0
  \end{cases}.
\end{align}
For the renormalized vector current, we use $G_E(0)=1$ and find nonzero results for the following combinations of $j$ and $\mu$:
\begin{align}
R_V^0 &= 1,
&\partial_1 R_V^2 &= -\frac{i}{2m}G_M(0), \label{eq1}\\
\partial_2 R_V^1 &=\frac{i}{2m} G_M(0),
&\partial_{1,2,3}^2 R_V^0 &= -\frac{1}{4m^2} - \frac{1}{3} r_E^2,
\label{eq2}
\end{align}
and for the axial current, 
\begin{align}
R_A^3 &= G_A(0),
&\partial_3 R_A^0 &= \frac{1}{2m} G_A(0),  \label{eq3}\\
\partial_{1,2}^2 R_A^3 &= -\frac{1}{4m^2} G_A(0) - \frac{1}{3} G_A(0) r_A^2,
&\partial_{3}^2R_A^3  &= -\frac{1}{4m^2} \left(G_A(0) + 2G_P(0)\right) - \frac{1}{3} G_A(0) r_A^2, \label{eq4}
\end{align}
with $\partial_j = \frac{\partial}{\partial p^j}$ and
\begin{align}
r_E^2 &= -\frac{6}{G_E(0)} \;\frac{dG_E}{dQ^2}\Big|_{Q^2=0}, \\
r_A^2 &= -\frac{6}{G_A(0)}\; \frac{dG_A}{dQ^2}\Big|_{Q^2=0}. 
\end{align} 
From Eq.~\eqref{eq1} and Eq.~\eqref{eq2}, we find the following relations for the nucleon magnetic moment $\mu = G_M(0)$ and squared charge radius $r_E^2$:
\begin{align}
\mu &= 2i\,m\,(R_{V}^2)',  \label{eq:mu}\\
r_E^2 &= -\frac{3}{4m^2} -3\;\frac{(R_V^0)''}{R_V^0},\label{eq:rE2}
\end{align}
where we average over equivalent vector components and directions:
\begin{align}
&(R_V^2)' = \frac{1}{2}(\partial_1R_V^2 - \nonumber \partial_2R_V^1),\\
&(R_V^0)'' = \frac{1}{3}(\partial_1^2R_V^0 + \partial_2^2R_V^0\; + \partial_3^2R_V^0).
\end{align} 
The squared axial radius $r_A^2$ and $G_P(0)$ can be evaluated using Eq.~\eqref{eq3} and Eq.~\eqref{eq4} as follows:
\begin{align}
r_A^2 & = -\frac{3}{4m^2} -\frac{3}{2} \frac{\partial_1^2 R_A^3  + \partial_2^2 R_A^3 }{R_A^3}, \label{eq:ra2} \\
G_P(0) &= m^2 \left( \partial_1^2 R_A^3  + \partial_2^2 R_A^3  - 2\partial_3^2 R_A^3 \right).\label{eq:GP_zero}
\end{align}

To estimate the excited-state effects contributing to the momentum derivatives of the ratio, we take the momentum derivatives of the leading contributions in Eq.~\eqref{rnra}, which leads to
\begin{equation}
\frac{\partial R}{\partial p_i} \Big|_{\vec p=0} \sim e^{-\Delta E_{10} T/2}, \qquad \frac{\partial^2 R}{\partial p_i^2} \Big|_{\vec p=0} \sim Te^{-\Delta E_{10} T/2}.
\end{equation}
Likewise, the expected excited-state effects in applying the summation method to the momentum derivatives of ratios are given by
\begin{equation}
\frac{\partial S}{\partial p_i} \Big|_{\vec p=0} \sim Te^{-\Delta E_{10} T}, \qquad \frac{\partial^2 S}{\partial p_i^2} \Big|_{\vec p=0} \sim T^2e^{-\Delta E_{10} T}.
\end{equation}

%------

%-------------------
\section{Lattice setup}
\label{LC}
%---------------------
We perform lattice QCD calculations using a tree-level Symanzik-improved gauge action~\cite{Durr:2010aw,Durr:2010vn} and 2+1 flavors of tree-level improved Wilson-clover quarks, which couple to the gauge links via two levels of HEX smearing. We carry out the calculations at the physical pion mass $m_\pi=135$ MeV, with lattice spacing $a=0.093$ fm, and a large volume $L_s^3 \times L_t= 64^4$ satisfying $m_\pi L = 4$.
We are measuring the isovector combination $u-d$ of the three-point functions, where the disconnected contributions cancel out. We renormalize the axial current using $Z_A$ from~\cite{Durr:2013goa} and the vector current by imposing $G_E^v(0)=1$.
Furthermore, we use three source-sink separations $T/a\in\{10, 13,16\}$ ranging from $0.9$ fm to $\sim 1.5$ fm, and we are using the summation method for removing contributions from excited states. We apply our analysis on $442$ gauge configurations, using all-mode-averaging~\cite{Blum:2012uh,Shintani:2014vja} with $64$ sources with approximate propagators and one source for bias correction per gauge configuration. For each source position we place nucleon sinks in both the forward and backward directions to double statistics and obtain a total of $56576$ samples.
We computed the momentum derivatives of the correlators only in the $x$ direction on a subset of the gauge configurations (75 configurations) and in the $x,y,$ and $z$ directions on the rest (367 configurations).
%----------
\section{Results}\label{sec5:results}
%---------
\subsection{Derivatives of the two-point functions}
\begin{figure}%[H]
\begin{center}
\includegraphics[width=0.45\textwidth]{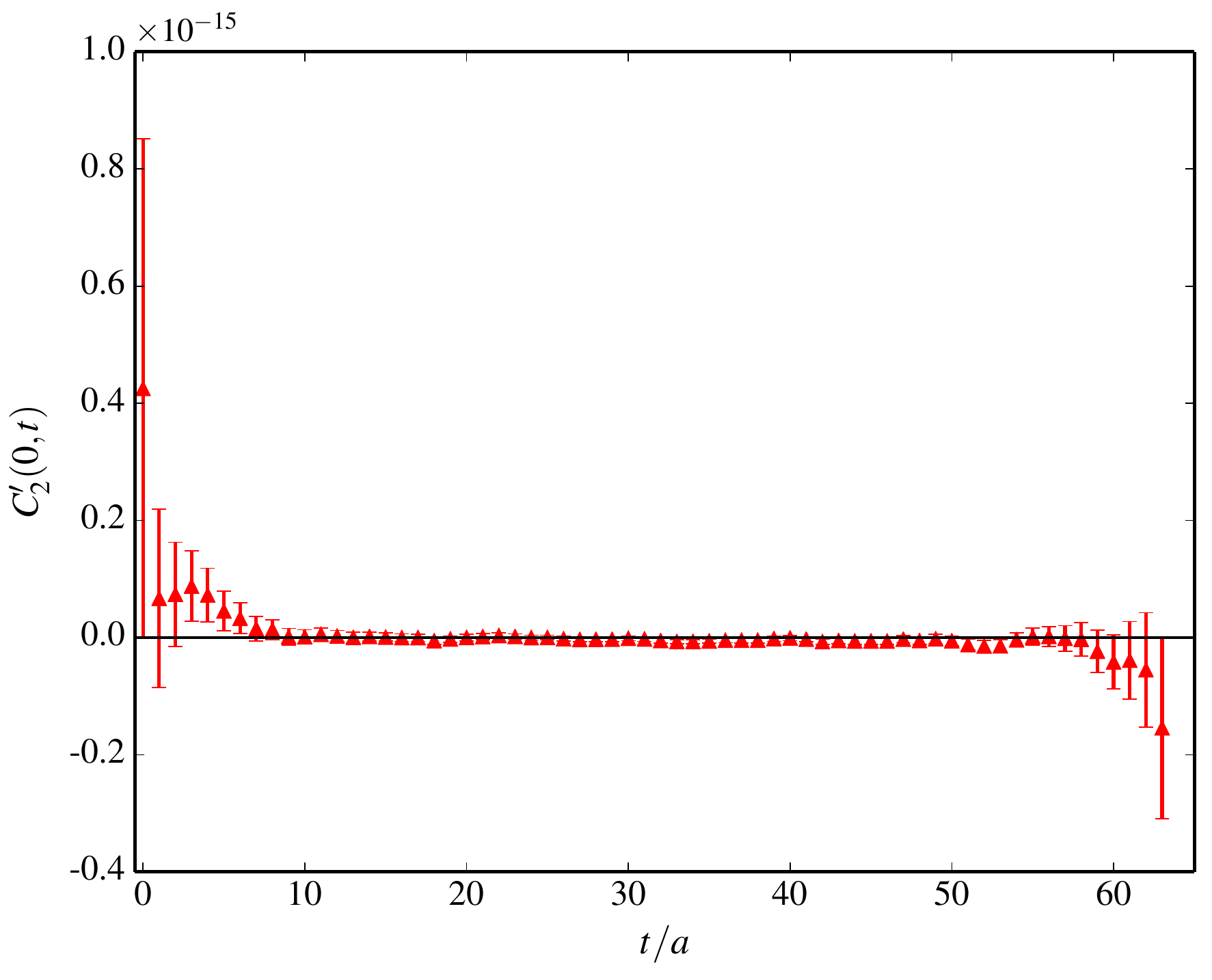}
\hspace{0.01\textwidth}
\includegraphics[width=0.45\textwidth]{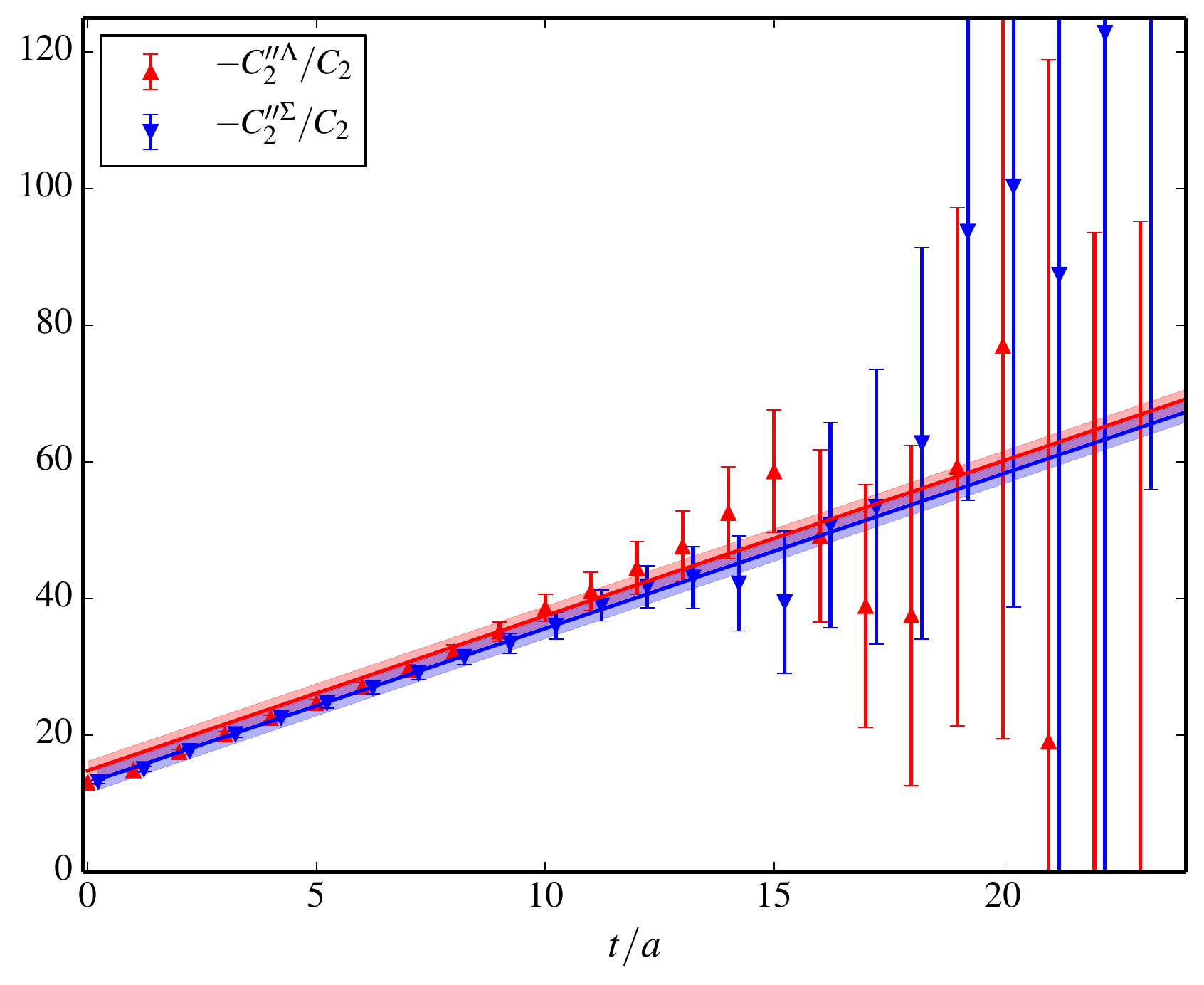}
\end{center}
  \caption{
  $C_2'(0,t)$ (left) and $-C_2''(0,t)^{\Lambda,\Sigma}/C_2(0,t)$ (right). The red and blue bands correspond to the combined fits of $C_2''(0,t)^{\Lambda,\Sigma}$ and $C_2(0,t)$.}
  \label{fig:ddc2}
\end{figure}

We begin by testing our method applied to the simpler case of two-point functions. From Eq.~\eqref{eq:gt-c2}, the ground-state contribution is
%Having $\pvec{p}'=0$, $\vec p=k \vec e_j$ and considering the case of an infinite temporal extent of the lattice, the ground-state contribution to the two-point function (using Eq.~\eqref{eq:gt-c2}) becomes
\begin{equation}
C_2(\vec{p},t)=\frac{Z(\vec p)^2 \left(E(\vec p)+m\right)}{E(\vec p)}e^{-E(\vec p)t}.
\end{equation}
The momentum derivatives of $C_2(\vec p, t)$ can then be evaluated at $\vec p=0$ and we obtain:
\begin{align} 
C_2(0,t)&= 2Z^2e^{-mt}, \label{eq:c20}\\
C_2'(0,t)&=4 Z Z'e^{-mt}, \\
C_2''(0,t)&=\frac{1}{m^2} \left[-(1+2mt)Z^2 + 4m^2 (Z')^2 + 4m^2 Z Z''\right]e^{-mt},\label{eq:d2c20}
\end{align}
where $Z\equiv Z(0)$.
We expect $C_2'(0,t)$ to vanish due to parity symmetry and our numerical results shown in the left part of Fig.~\ref{fig:ddc2} confirm that, which allows us to set $Z'(0)=0$ in Eq.~\eqref{eq:d2c20}. 
We apply a combined 1-state fit for $C_2(0,t)$ and $C_2''(0,t)^{\Lambda,\Sigma}$ using Eq.~\eqref{eq:c20} and Eq.~\eqref{eq:d2c20} with $Z,Z''$ and $m$ being the fit parameters. The results of these fits are shown in Fig.~\ref{fig:ddc2}, where the slight differences between the momentum derivatives of $\Sigma_r$ and $\Lambda_r$ two-point functions give an indication of the systematic errors associated with the derivative method and motivate the approach described in Sec.~\ref{flavourstructure} for isolating $\Sigma_r \to N$ from  $\Lambda_r \to N$ three point functions when extracting the momentum derivatives of the matrix elements.
 
\begin{figure}
\centering
\includegraphics[width=0.5\textwidth]{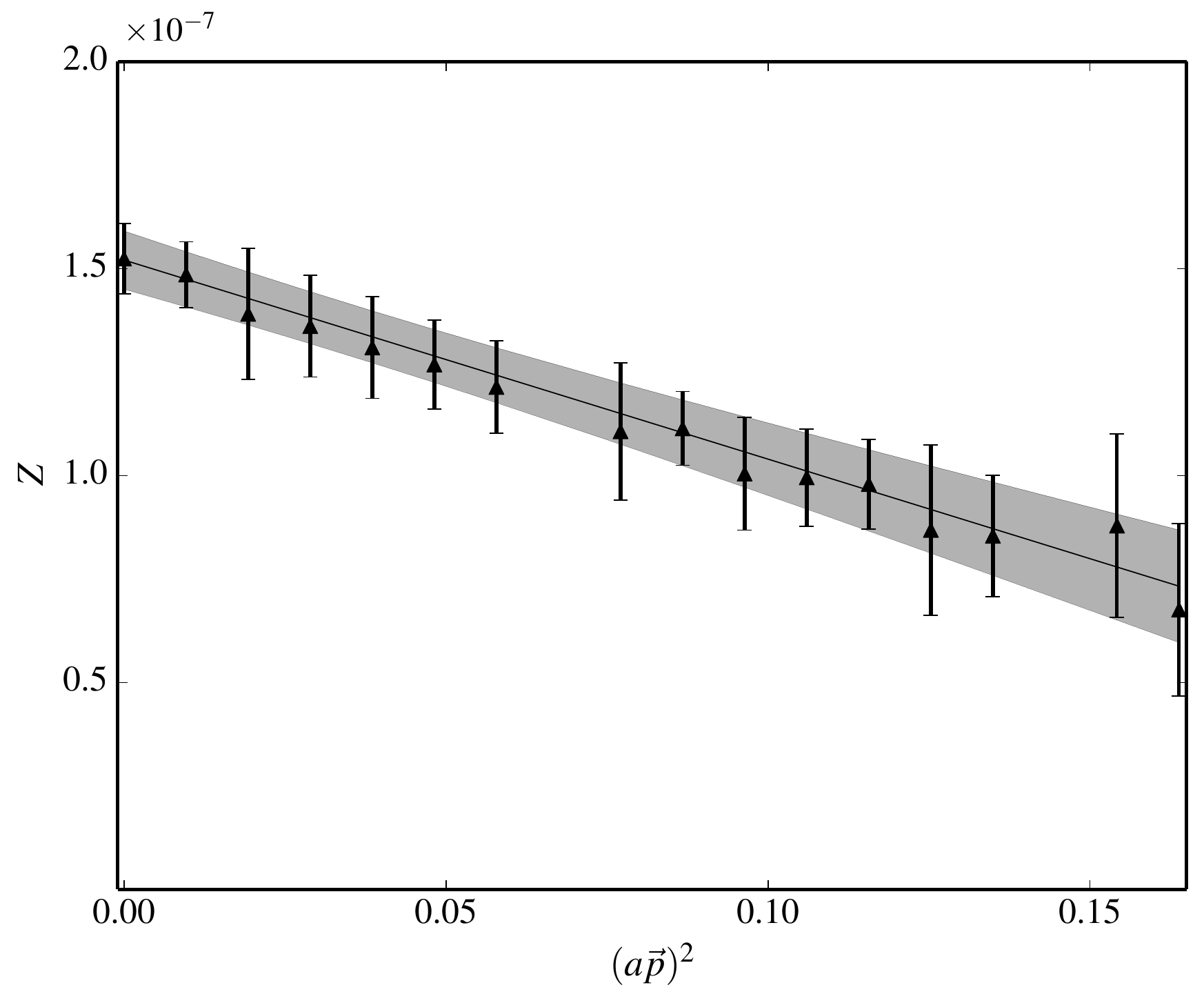}
\caption{The derived values for $Z(\vec p^{\,2})$ from two-state fits of $C_2(\vec p,t)$ (black points) followed by a linear fit (grey band) for extracting $Z(0)$ and $Z''(0)$.}
\label{fig:Z_2state}
\end{figure}

We also try another approach for extracting $Z(0)$ and $Z''(0)$ where we apply two-state fits to $C_2(\vec p,t)$ for different discrete values of $\vec p^{\,2}$ which allows us to extract $Z(\vec p^{\,2})$. The extracted values for $Z(\vec p^{\,2})$ are consistent with a linear dependence on $(a\vec p)^2$.
By applying a linear fit to $Z(\vec p^{\,2})$ against ${\vec p}^{\,2}$, $Z(0)$ can be obtained from the intercept and $Z''(0)$ from the slope as $Z''(0) = 2\frac{\partial Z(\vec p^{\,2})}{\partial \vec p^{\,2}}$. This is shown in Fig.~\ref{fig:Z_2state}.

Table~\ref{tab:ZZp} reports a comparison between the extracted values for $Z(0)$ and $Z''(0)$ using the two different approaches and when using $[C_2''(0,t)]^\Sigma$ and $[C_2''(0,t)]^\Lambda$ in the combined fit. All fit methods lead to consistent values for both $Z(0)$ and $Z''(0)$.
% However,
%we obtain different values for $Z''(0)$ when using $[C_2''(0,t)]^\Lambda$ or $[C_2''(0,t)]^\Sigma$ in the combined fit. The values for $Z''(0)$ obtained from the fit to $Z(\vec p^{\,2})$ and the combined fit when using $[C_2''(0,t)]^\Lambda$ are consistent. 
%with consistency between $Z''(0)$ obtained from the fit to $Z(\vec p^{\,2})$ and the combined fit when using $[C_2''(0,t)]^\Lambda$.
%, $Z''(0)$ value obtained from the combined fit using $[C_2''(0,t)]^\Sigma$ is not consistent with its value from the combined fit using $[C_2''(0,t)]^\Lambda$ and the linear fit approach.

\begin{table}%[H]
\centering
\renewcommand{\arraystretch}{1.5}
\begin{tabular}{|P{4cm}|P{3.5cm}|P{3.5cm}|}
    \hline
    Method &$Z(0) \times 10^{7}$ & $Z''(0) \times 10^{7}$ \\ 
    \hline
     Fit $C_2(0,t)$ and $C_2''(0,t)^\Lambda$ & $1.633(14)$ &$-9.9(1.1)$ \\
     \hline
     Fit $C_2(0,t)$ and $C_2''(0,t)^\Sigma$ & $1.635(15)$ &$-8.9(1.2)$ \\
     \hline
     Fit $Z(\vec p^{\,2})$ & $1.521(70)$ & $-9.6(1.8)$ \\ 
     \hline
\end{tabular}
\caption{Resulting values for $Z(0)$ and $Z''(0)$ using either the combined fit of $C_2(0,t)$ and $C_2''(0,t)^{\Lambda,\Sigma}$ or the fit to $Z(\vec p^{\,2})$.}
\label{tab:ZZp}
\end{table}

%---------
\subsection{Electromagnetic form factors}
\begin{figure}%[H]
\begin{center}
\includegraphics[width=0.45\textwidth]{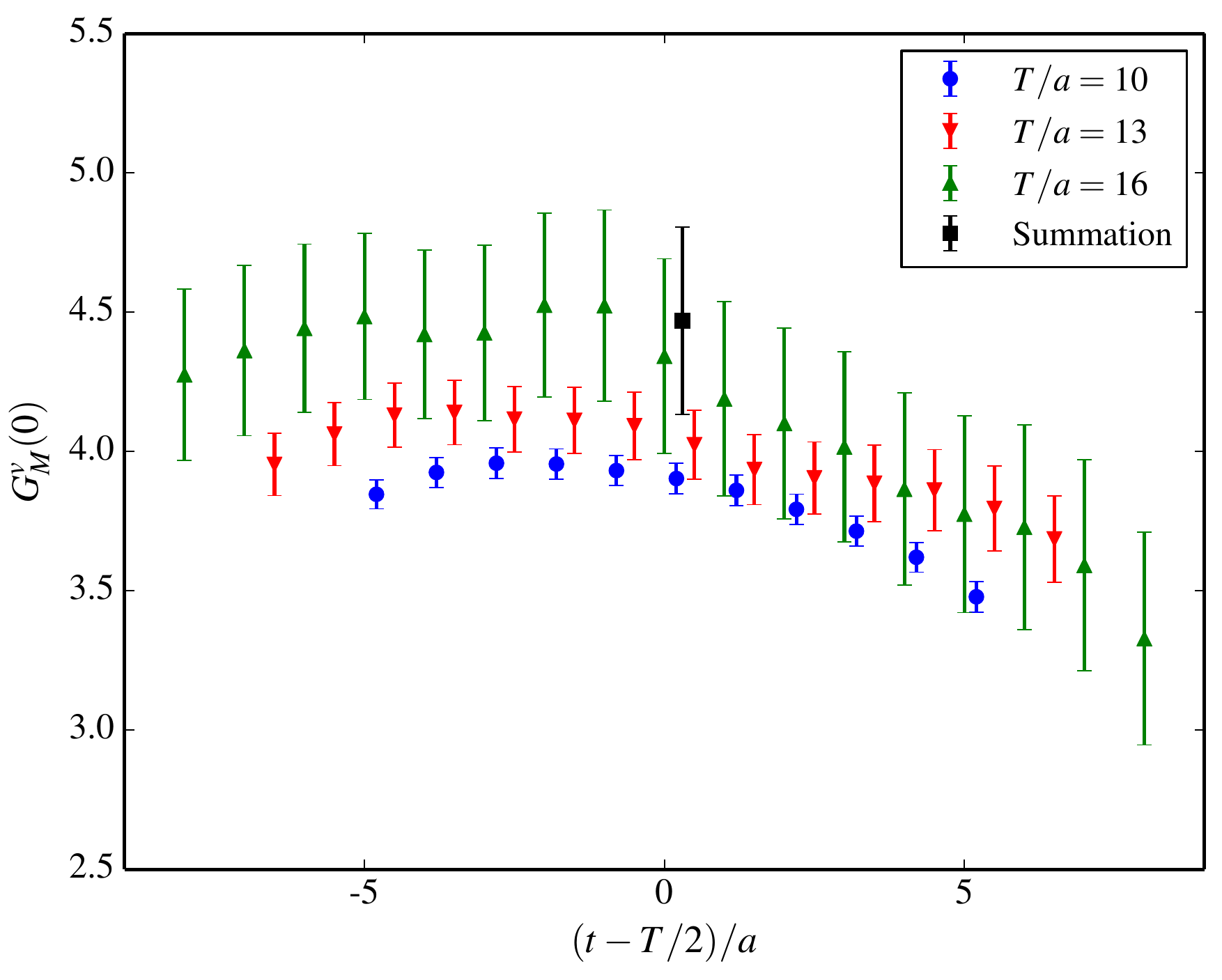}
\hspace{0.01\textwidth}
\includegraphics[width=0.45\textwidth]{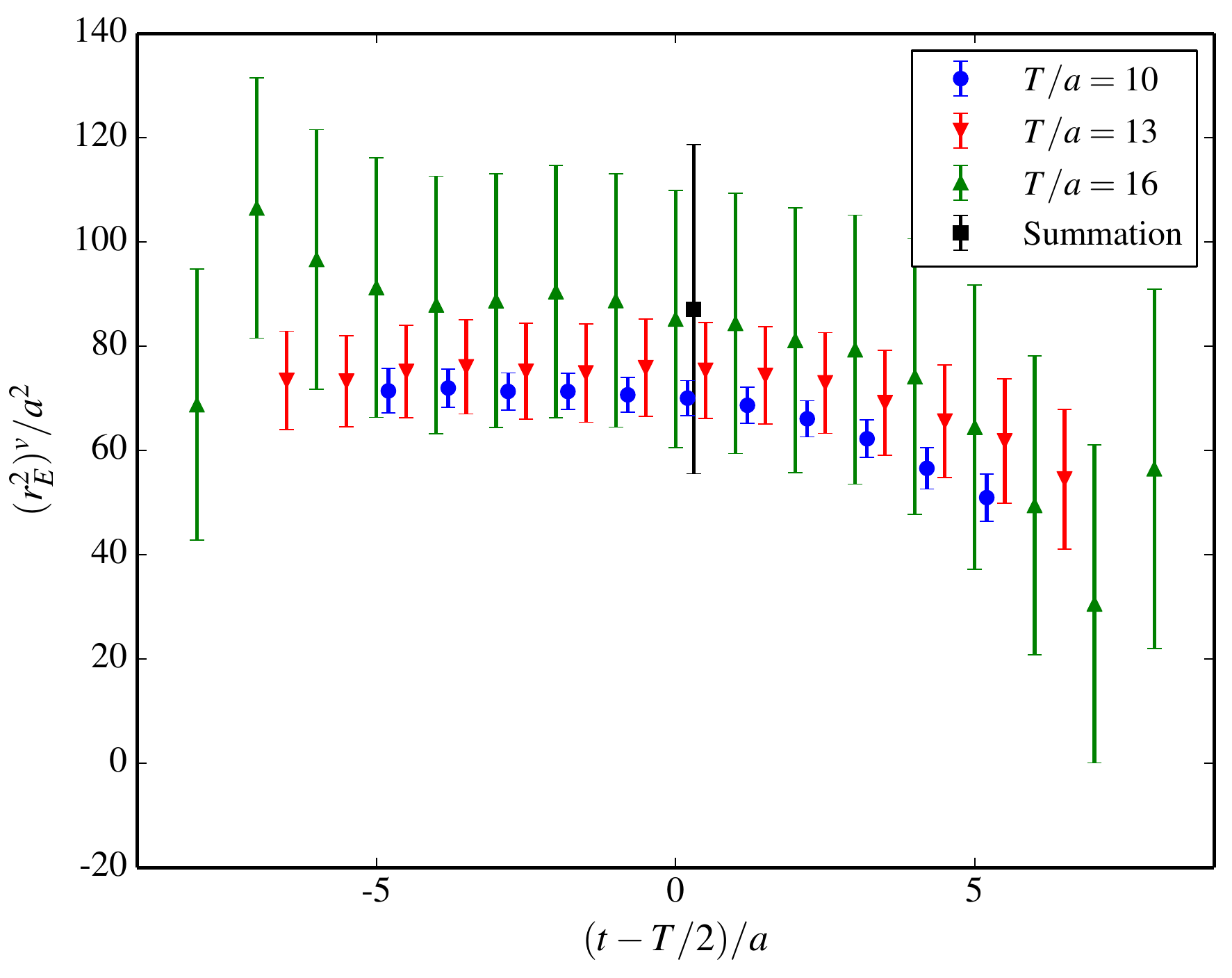}
 \end{center}
  \caption{Isovector magnetic moment (left) and isovector charge radius (right). For both $\mu^v$ and $(r_E^2)^v/a^2$, results from the ratio method are shown using source-sink separations $T/a \in \{10, 13, 16\}$, as well as the summation method.}
  \label{fig1}
\end{figure}

The ``plateau plots'' in Fig.~\ref{fig1} show the results we obtain using the momentum derivative approach for both $G_M^{v}(0)$ (left), computed using Eq.~\eqref{eq:mu}, and $(r_E^2)^v/a^2$ (right), extracted from Eq.~\eqref{eq:rE2}. In each case, we show results from both the ratio method and the summation method. $G_M^{v}(0)$ increases for increased source-sink separations, indicating that the excited-state contributions are significant in this case. The relative statistical uncertainty is much larger for $(r_E^2)^v/a^2$, and therefore we are unable to resolve any significant excited-state effects.

\begin{figure}%[H]
\begin{center}
\includegraphics[width=0.9\textwidth]{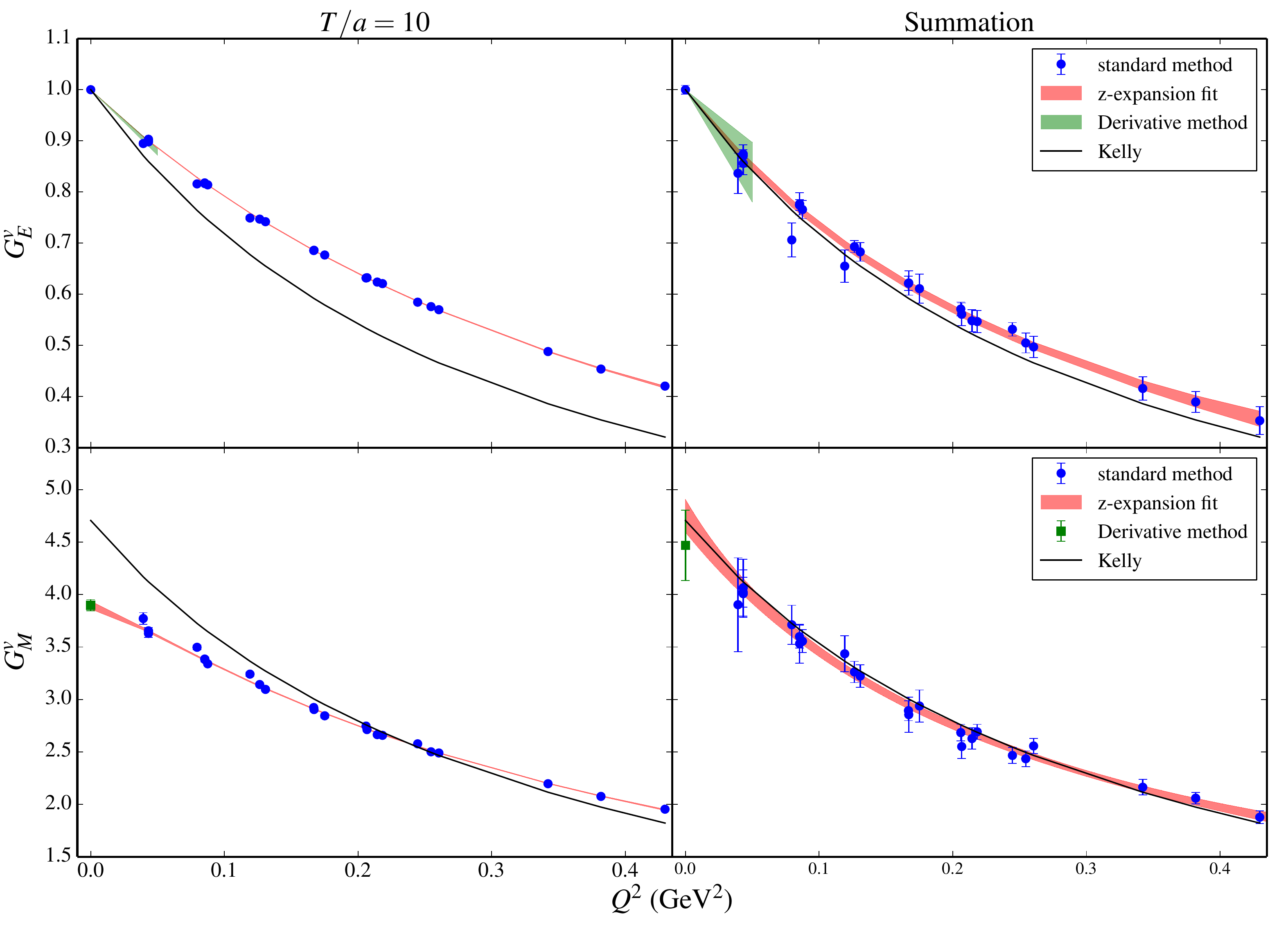}
\end{center}
  \caption{Isovector electric (top row) and magnetic (bottom row) form factors using both the ratio method with $T=10\;a$ (left column) and the summation method (right column). The blue points show results from the standard method and the red bands show a $z$-expansion fit to those points. The green band (top) and point (bottom) show the slope and value of the respective form factor at $Q^2=0$, computed using
the momentum derivative method. The black curves result from a phenomenological fit to
experimental data by Kelly~\cite{Kelly:2004hm}.}
\label{fig2}
\end{figure}

Figure~\ref{fig2} shows a comparison between our results using the derivative method and the traditional approach for both the isovector magnetic moment $\mu^{v}=G_M^{v}(0)$ (bottom row) and the isovector charge radius $(r_E^2)^v$ (top row). In Fig~\ref{fig2}, we present the results extracted using both the ratio method with the smallest source-sink separation $T/a=10$ and the summation method. When going to the summation method, $G_E^v(Q^2)$ decreases significantly whereas $G_M^v(Q^2)$ increases (especially for small $Q^2$) towards the corresponding phenomenological curve from Kelly~\cite{Kelly:2004hm}. This shows the non-trivial contribution from excited states associated with the ratio method using $T/a=10$. The summation points for $G_E^v(Q^2)$ lie slightly above the corresponding Kelly curve while those for $G_M^v(Q^2)$ show a good agreement with the Kelly curve. The derivative method's results for both $G_M^{v}(0)$ and $(r_E^2)^v$ using the summation method are consistent with both the traditional method's results and the experiment but with statistical errors roughly twice as large as the traditional approach for the isovector magnetic moment and three times as large for the isovector charge radius, as reported in Table~\ref{tab:ff}.

\subsection{Axial form factors}
\begin{figure}%[H]
%\centering
\begin{center}
\includegraphics[width=0.45\textwidth]{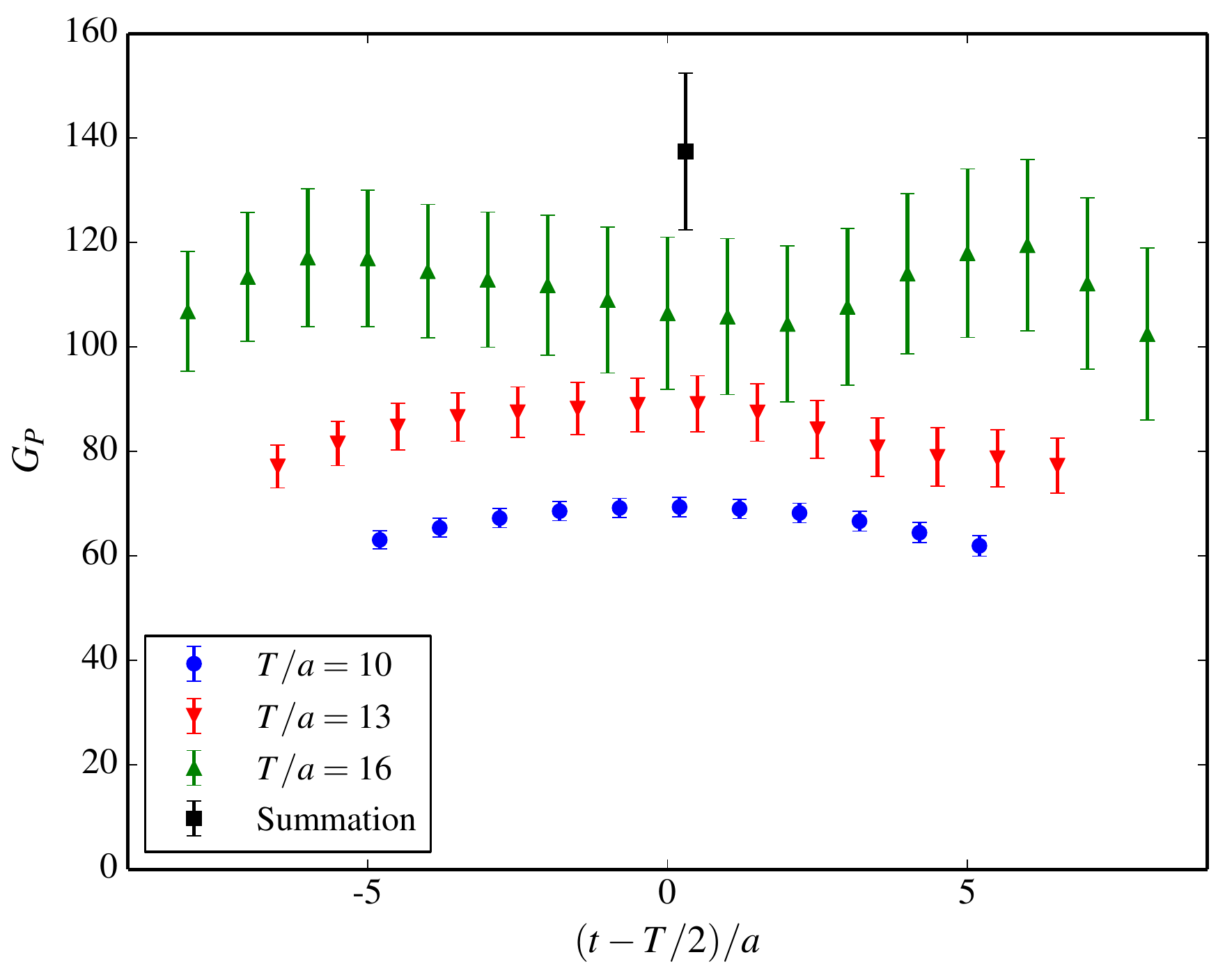}
\hspace{0.01\textwidth}
\includegraphics[width=0.45\textwidth]{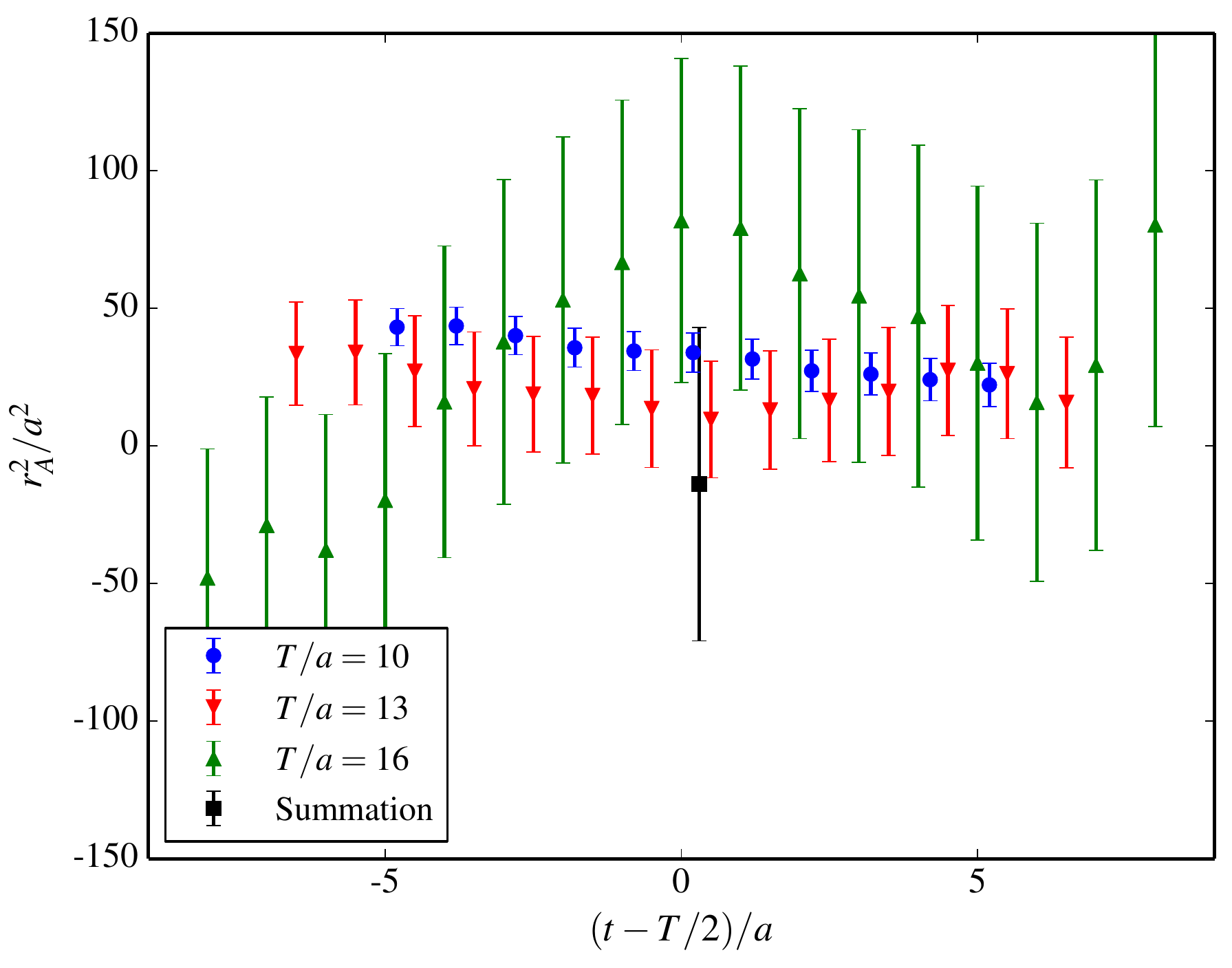}
\end{center}
  \caption{ The induced pseudoscalar form factor at $Q^2=0$ (left) and nucleon axial radius (right). For both $G_P(0)$ and $r_A^2/a^2$, results from ratio method are shown using source-sink separations $T/a \in \{10, 13, 16\}$, as well as the summation method.}
  \label{fig3}
\end{figure}

The left-hand side of Fig.~\ref{fig3} shows the isovector induced pseudoscalar form factor $G_P^v(0)$ extracted using the derivative method, Eq.~\eqref{eq:GP_zero}. The right-hand side of the same figure shows the extracted $r_A^2$ using Eq.~\eqref{eq:ra2}. Figure~\ref{fig3} shows the plateau plots for both quantities corresponding to the three available source-sink separations in addition to the summation points. 
For $G_P^v(0)$, we see a large increase with the source-sink separation, indicating substantial 
excited-state effects, and that leads us to conclude that the summation point may not be free from excited-state effects. For $r_A^2$, the statistical errors are too large to detect any excited-state effects.

\begin{figure}%[H]
\begin{center}
\includegraphics[width=0.9\textwidth]{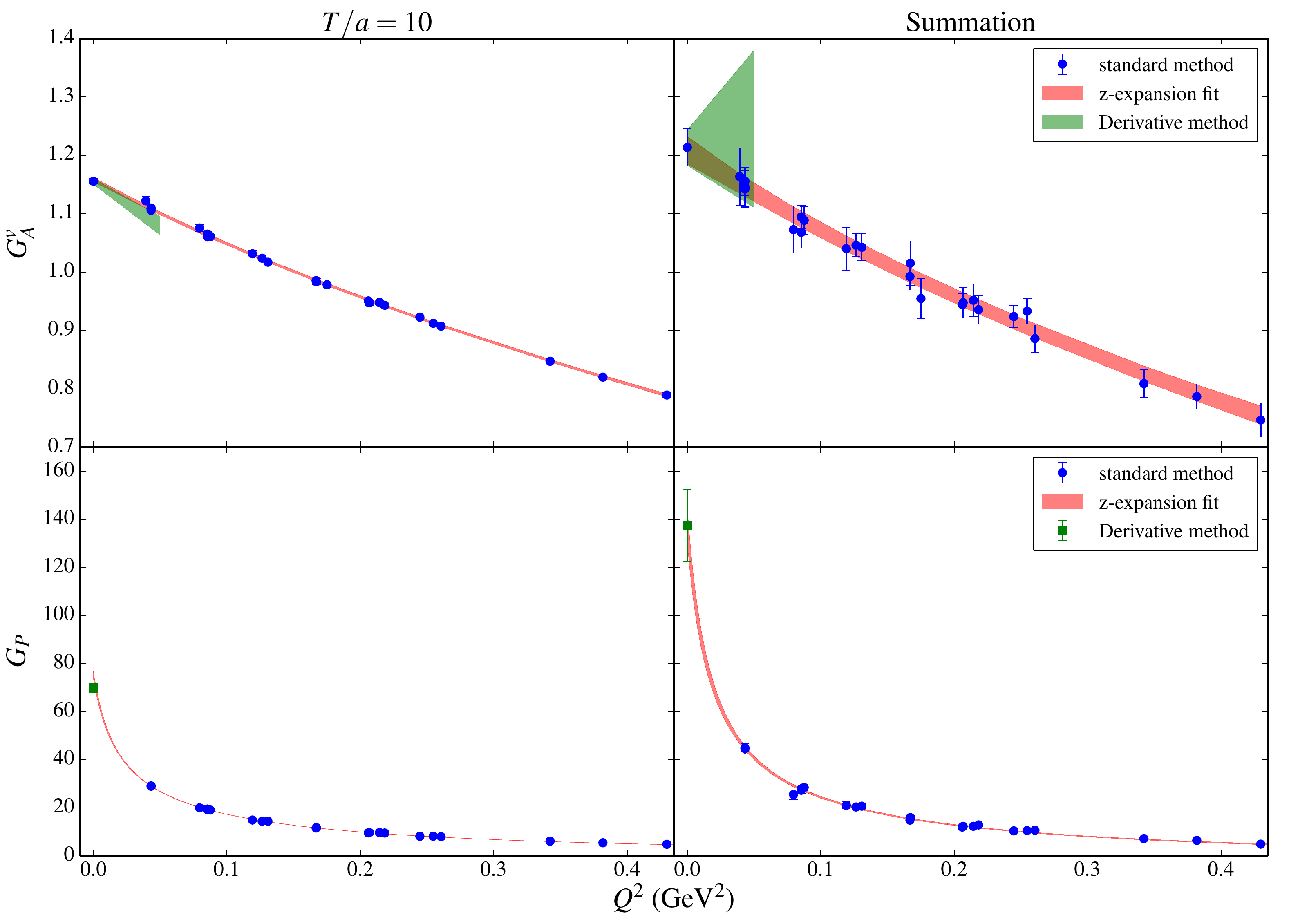}
\end{center}
  \caption{Nucleon axial (top row) and induced pseudoscalar (bottom row) form factors using both the ratio method for $T=10\;a$ (left column) and the summation method (right column). The blue points show results from
the standard method and the red bands show a $z$-expansion fit to those points. The green band (top) and
point (bottom) show the slope and value of the respective form factor at $Q^2=0$, computed using
the momentum derivative method.}
\label{fig4}
\end{figure}

A comparison between our results using the derivative method and the traditional method for both $r_A^2$ and $G_P^v(0)$ is shown in Fig.~\ref{fig4}, top and bottom row, respectively. Shown are results from both the ratio method with $T/a=10$ and the summation method. 
Both $G_A^v(Q^2)$ and $G_P^v(Q^2)$ increase when going to the summation method indicating the significant excited-state contributions for the ratio method with $T/a=10$. 
The extracted value for the axial radius using the derivative method has a much larger statistical error compared to its value from the traditional approach. For $G_P^v$ in Fig.~\ref{fig4}, before fitting we remove the pion pole that is present in the form factor, and then restore it in the final fit curve as was discussed in Sec.~\ref{sec:2}. At $T/a=10$, there is a significant disagreement between $G_P(0)$ from the traditional and the derivative approaches which is likely due to excited-state effects. The value for $G_P^v(0)$ using the summation method and the derivative approach seems to be in good agreement with its value from the traditional approach despite the large extrapolation caused by the inclusion of the pion pole in the fit. However, $G_P^v(0)$ obtained from the derivative method has statistical uncertainties roughly twice as large as the traditional approach. Our results for the axial form factors are reported in Table~\ref{tab:ff}.     


\begin{table}%[H]%!htbp
\centering
\makebox[\textwidth][c]{
\begin{tabular}{c|cc|cc|cc|cc}
\toprule
 \multirow{2}{*}{}&\multicolumn{2}{c|}{$\mu^v$} & \multicolumn{2}{c|}{$(r_E^2)^v$ [fm]$^2$} & \multicolumn{2}{c|}{$G_P(0)$} & \multicolumn{2}{c}{$r_A^2$ [fm]$^2$}\\
 \cline{2-9}
   &$T/a=10$& Summation& $T/a=10$& Summation &$T/a=10$& Summation &$T/a=10$& Summation \\
\cline{2-9}
%\hline
Traditional method  &3.899(38)&4.75(15) &0.608(15)& 0.787(87)&75(1)& 137(7)&0.249(12)&0.295(68)\\
 Derivative method &3.898(54)&4.46(33)&0.603(29)&0.753(273)&69(1)& 137(15)&0.288(61)&$-$0.120(492)\\ 
\bottomrule
\end{tabular}
}
\caption{Numerical results for the four different nucleon observables at $Q^2=0$, computed with the traditional method (via $z$ expansion fit to the form factor shape) and with the derivative method.}
\label{tab:ff}
\end{table}

\iffalse
\begin{table}%[H]
\centering
\renewcommand{\arraystretch}{1.5}
\begin{tabular}{|P{3cm}|P{1.5cm}|P{2cm}|P{1.5cm}|P{1.5cm}|P{1.5cm}|P{1.5cm}|P{1.5cm}|P{1.5cm}|}
\hline
\text{Quantity}&$\mu^v$& $(r_E^2)^v$ [fm]$^2$ & $G_p(0)$ & $r_A^2$ [fm]$^2$ & & & &\\
\hline
Traditional method& 4.75(15)& 0.787(87) & 137(7) & 0.295(68)\\
\hline
Derivative method & 4.46(33)& 0.69(28) & 137(15) & -0.12(51)\\
\hline
\end{tabular}
\caption{A comparison between the derivative method and the traditional one}
\label{tab:ff}
\end{table}
\fi
%-----------
\section{Summary and outlook}\label{sec6:conclusion}
%--------------
In this paper, we presented a derivative method for computing nucleon observables at zero momentum transfer. Using this method helps to avoid the model dependence and the large extrapolation needed in the traditional approach for computing such quantities.
We applied the derivative method to the nucleon isovector magnetic moment and electric charge radius as well as the isovector induced pseudoscalar form factor at $Q^2=0$ and the axial radius.

We confirm that our approach produces results consistent with those obtained using the traditional method. For $G_M(0)$ and $G_P(0)$, there is excellent agreement between the two approaches. This is particularly remarkable in the latter case, since the pion pole produces a very large effect in the extrapolation of $G_P(Q^2)$ to $Q^2=0$.
However, we found that this approach suffers from large statistical uncertainties, especially for the isovector charge and axial radii. This may be connected with the fact that these quantities require
a second momentum derivative. However, $G_P(0)$ also requires two derivatives and is not as noisy.
Our quoted errors are statistical; we still need to estimate and improve control over systematic uncertainties in order to have a reliable calculation.
% We still need to improve the control over uncertainties in order to have a reliable calculation.
 The difference between the CODATA value of $(r_E^2)^v$ and its muonic 
hydrogen measurement is $\sim 0.06$ fm$^2$, so it will be a challenge to calculate the charge radius with a total uncertainty significantly less than that. 

Our present setup of the derivative method includes computing the momentum derivatives of the nucleon correlators with respect to only the initial nucleon momentum. As suggested originally for the pion charge radius in Ref.~\cite{Tiburzi:2014yra}, one can  alternatively obtain the radius by computing the mixed-momentum derivatives of three-point functions i.e., first-order momentum derivatives with respect to both initial- and final-state momenta. A calculation including this alternative approach is currently underway; preliminary results 
suggest that the statistical uncertainty for the radii is significantly reduced~\cite{hasan2017}.

\acknowledgments
We thank the Budapest-Marseille-Wuppertal collaboration for making their configurations available to us. This research used resources at Forschungszentrum J\"lich and on CRAY XC40 (Hazel Hen) at HLRS. JG was supported in part by the PRISMA Cluster of Excellence at the University of Mainz, and SM is supported in part by NSF grant PHY-1520996. SM and
SS are supported by the RIKEN BNL Research Center under its joint tenure track fellowships with the University of Arizona and Stony Brook University, respectively. ME, JN, and AP are supported in part by the
Office of Nuclear Physics of the U.S. Department of Energy (DOE) under grants DE-FG02-96ER40965, DE-SC-0011090, and DE-FC02-06ER41444,respectively. SK and NH received support from Deutsche Forschungsgemeinschaft grant SFB-TRR 55.
\\
Calculations for this project were done using the Qlua software suite.

\bibliography{DM_2017}
\bibliographystyle{utphys-noitalics}

\end{document}